\newif\ifisarxiv

\isarxivtrue

\ifisarxiv
    \newcommand{\fullver}[1]{#1}
    \newcommand{\confver}[1]{}
\else
    \newcommand{\fullver}[1]{}
    \newcommand{\confver}[1]{#1}
\fi

\ifisarxiv

\documentclass[10pt]{article}

\usepackage[margin=1in]{geometry}
\usepackage[numbers, compress]{natbib}

\title{A Dataset on Malicious Paper Bidding in Peer Review}

\author{
  Steven Jecmen \\
  Carnegie Mellon University\\
  \texttt{sjecmen@cs.cmu.edu}\\
  \and
    Minji Yoon \\
  Carnegie Mellon University\\
  \texttt{minjiy@cs.cmu.edu} \\
  \and
  Vincent Conitzer \\
  Carnegie Mellon University\\
  \texttt{conitzer@cs.cmu.edu} \\
  \and
  Nihar B. Shah \\
  Carnegie Mellon University\\
  \texttt{nihars@cs.cmu.edu} \\
  \and
  Fei Fang \\
  Carnegie Mellon University\\
  \texttt{feifang@cmu.edu} \\
}
\author{
  \begin{tabular}{ c c c }
       Steven Jecmen &    Minji Yoon   &    Vincent Conitzer \\
    Carnegie Mellon University & Carnegie Mellon University & Carnegie Mellon University \\
    \texttt{sjecmen@cs.cmu.edu} & \texttt{minjiy@cs.cmu.edu} & \texttt{conitzer@cs.cmu.edu} \\
  \end{tabular} \\ \\
  \begin{tabular}{ c c }
       Nihar B. Shah   &    Fei Fang   \\
    Carnegie Mellon University & Carnegie Mellon University \\
    \texttt{nihars@cs.cmu.edu} & \texttt{feifang@cmu.edu} \\
  \end{tabular} 
}
\date{}
\date{}

\else 
\documentclass[sigconf]{acmart}

\copyrightyear{2023}
\acmYear{2023}
\setcopyright{rightsretained}
\acmConference[WWW '23]{Proceedings of the ACM Web Conference 2023}{May 1--5, 2023}{Austin, TX, USA}
\acmBooktitle{Proceedings of the ACM Web Conference 2023 (WWW '23), May 1--5, 2023, Austin, TX, USA}
\acmDOI{10.1145/3543507.3583424}
\acmISBN{978-1-4503-9416-1/23/04}

\title{A Dataset on Malicious Paper Bidding in Peer Review}



\author{Steven Jecmen}
\affiliation{%
  \institution{Carnegie Mellon University}
  \city{Pittsburgh}
  \state{PA}
  \country{USA}}
\email{sjecmen@cs.cmu.edu}
\author{Minji Yoon}
\affiliation{%
  \institution{Carnegie Mellon University}
  \city{Pittsburgh}
  \state{PA}
  \country{USA}}
\email{minjiy@cs.cmu.edu}
\author{Vincent Conitzer}
\affiliation{%
  \institution{Carnegie Mellon University}
  \city{Pittsburgh}
  \state{PA}
  \country{USA}}
\email{conitzer@cs.cmu.edu}
\author{Nihar B. Shah}
\affiliation{%
  \institution{Carnegie Mellon University}
  \city{Pittsburgh}
  \state{PA}
  \country{USA}}
\email{nihars@cs.cmu.edu}
\author{Fei Fang}
\affiliation{%
  \institution{Carnegie Mellon University}
  \city{Pittsburgh}
  \state{PA}
  \country{USA}}
\email{feifang@cmu.edu}

\fi

\usepackage[utf8]{inputenc} 
\usepackage[T1]{fontenc}    
\PassOptionsToPackage{hyphens}{url}
\usepackage{hyperref}       
\usepackage{url}            
\usepackage{booktabs}       
\usepackage{amsfonts}       
\usepackage{nicefrac}       
\usepackage{microtype}      
\usepackage{xcolor}         

\usepackage{comment}
\usepackage{subcaption}
\usepackage{graphicx}
\usepackage{enumitem}
\usepackage{wrapfig}
\usepackage{amsmath}

\newcommand{\adrev}{r}
\newcommand{\adpap}{p}
\newcommand{\simmat}{S}
\newcommand{\areamat}{A}
\newcommand{\bidmat}{B}

\begin{document}

\fullver{\maketitle}

\begin{abstract} 
In conference peer review, reviewers are often asked to provide ``bids'' on each submitted paper that express their interest in reviewing that paper. A paper assignment algorithm then uses these bids (along with other data) to compute a high-quality assignment of reviewers to papers. However, this process has been exploited by malicious reviewers who strategically bid in order to unethically manipulate the paper assignment, crucially undermining the peer review process. For example, these reviewers may aim to get assigned to a friend's paper as part of a quid-pro-quo deal. A critical impediment towards creating and evaluating methods to mitigate this issue is the lack of any publicly-available data on malicious paper bidding. In this work, we collect and publicly release a novel dataset to fill this gap, collected from a mock conference activity where participants were instructed to bid either honestly or maliciously. We further provide a descriptive analysis of the bidding behavior, including our categorization of different strategies employed by participants. Finally, we evaluate the ability of each strategy to manipulate the assignment, and also evaluate the performance of some simple algorithms meant to detect malicious bidding. The performance of these detection algorithms can be taken as a baseline for future research on detecting malicious bidding. 
\end{abstract} 

\confver{\maketitle}

\section{Introduction}
\confver{
Academic peer review is a widely-used application of human computation,  involving inputs from hundreds or thousands of reviewers around the world. Various issues in peer review have recently received attention from the human computation research community~\cite{jecmen2021nearoptimal,dhull2022price}. 
}
In peer review, expert reviewers must be assigned to papers which they are qualified for and interested in reviewing. Large scientific conferences typically assign reviewers by first computing a similarity score for each reviewer-paper pair, which indicates the expected quality of review from that reviewer for that paper. These similarity scores are computed from various components~\cite{shah2021survey}, including the paper and reviewer subject areas, text-matching with the reviewer's past work~\cite{mimno07topicbased,liu14graphpropagation,rodriguez08coauthorsip,tran17expertsuggestion,charlin13tpms}, and reviewer ``bids'' on each paper. After computing similarities, the conference determines the assignment by solving an optimization problem, where the objective is to maximize total similarity subject to constraints that each paper is assigned the desired number of reviewers, each reviewer is assigned to (at most) the desired number of papers, and no reviewer is assigned to a paper with which they have declared a conflict of interest~\cite{charlin13tpms,Long13gooadandfair,goldsmith07aiconf,tang10constraied,flach2010kdd,taylor08assignment}.

Paper bidding is a major component of the similarity computation, in which each reviewer indicates their level of interest in reviewing each paper from a list of options (e.g., ``Not willing'', ``In a pinch'', ``Willing'', ``Eager''). These bids are often given a high weight in the similarity computation, allowing each reviewer to have a significant level of control over their own assignment. For example, at the AAAI 2021 conference~\citep{leytonbrown2022matching}: ``{\it Reviewers were assigned papers for which they bid positively (willing or eager) 77.4\% of the time. A back-of-the-envelope calculation leads us to estimate that 79.3\% of these matches may not have happened had the reviewer not bid positively.}'' This is desirable since paper bidding allows each reviewer to express their own expertise in a flexible manner and helps to ensure that reviewers will not be asked to review papers in which they have no interest.

However, heavy reliance on paper bidding opens up the peer review process to be exploited by malicious reviewers. These reviewers attempt to manipulate the paper assignment through bidding, with the aim of getting assigned to some ``target papers'' and leaving them a dishonest review. Moreover, groups of reviewers may unethically collude (after failing to report conflicts of interest), working together to get assigned to review each other's papers and leave positive reviews. 
Examples of this form of collusion have been discovered in multiple computer science conferences~\cite{Vijaykumar_2020,littman2021collusion}:
``{\it The colluders hide conflicts of interest, then bid to review these papers, sometimes from duplicate accounts, in an attempt to be assigned to these papers as reviewers.}'' This sort of manipulation of the paper assignment critically undermines the peer review process and erodes the scientific community's trust in its effectiveness. Thus, addressing this unethical behavior is an urgent challenge for modern conferences. 

In recent years, approaches to addressing bid manipulation have been proposed in the research literature~\cite{jecmen2020manipulation,wu2021making} and implemented by conferences~\cite{shah2021survey,leytonbrown2022matching}. However, a significant challenge in solving this problem is that there is no dataset on which proposed algorithms can be evaluated and compared. Data from real conferences often cannot be released due to privacy concerns. More importantly, it is nearly impossible to claim for sure which reviewers were acting maliciously. Any public information about the aforementioned incidents of collusion that have been uncovered is kept vague. Researchers thus must rely on synthetic implementations of malicious behavior in order to test their algorithms, without any hard data on which to base the implementation. Such data is necessary in order to develop effective solutions to this important issue, despite the impossibility of collecting the ideal real-world dataset.  

\paragraph{Our contributions:} 
\ifisarxiv \begin{enumerate}[leftmargin=*] \fi 
\fullver{\item}\confver{(1)} We construct and publicly release a dataset containing bidding data from human participants in a mock conference setting (Section~\ref{sec:data}), taking the first step towards filling a critical gap in the research landscape on this important problem. To our knowledge, this is the first dataset of this kind available to other researchers. Participants were instructed to bid first as an honest reviewer and then as a malicious reviewer, so the data contains both honest bids and malicious bids with ground-truth labels on whether the behavior was malicious. 
\fullver{\item}\confver{(2)} We supplement this dataset with descriptions of participants' behavior and a categorization of the strategies employed to manipulate the assignment (Section~\ref{sec:desc}).
\fullver{\item}\confver{(3)} We empirically evaluate the success of participant strategies in terms of their ability to manipulate the assignment (Section~\ref{sec:successeval}). 
\fullver{\item}\confver{(4)} We propose several simple detection algorithms, which can serve as baselines for other researchers aiming to develop algorithms to detect bid manipulation. We then evaluate the success of these algorithms at detecting different strategies (Section~\ref{sec:detecteval}). 
\fullver{\item}\confver{(5)} We synthetically scale up the dataset and provide additional large-scale evaluations (Section~\ref{sec:synth}). 
\ifisarxiv \end{enumerate} \fi 
The dataset and our analysis code is publicly available at \url{https://github.com/sjecmen/malicious_bidding_dataset}.

\section{Related Work} \label{sec:relwork}
The problem of malicious bid manipulation has been taken seriously by recent conferences, who have attempted a variety of approaches to address it. For example, AAAI 2021 implemented several techniques to break up colluding reviewers, including preventing two-cycles in the reviewer assignment and requiring geographic diversity among the reviewers assigned to the same paper~\cite{leytonbrown2022matching}. They also required a minimum number of positive bids from each reviewer, with the aim of preventing reviewers from targeting a specific paper. AAAI 2022 similarly implemented a geographic diversity constraint and a minimum number of positive bids~\cite{shah2021survey}. Conferences may also use techniques to flag some reviewers for later manual investigation. 

Other approaches to addressing this problem have been proposed in the research literature. Jecmen et al.~\cite{jecmen2020manipulation} propose randomizing the reviewer assignment by limiting the probability that any reviewer can be assigned to any paper, thus limiting the probability with which a malicious reviewer can be assigned to a target paper. This technique was used for part of the AAAI 2022 assignment~\cite{shah2021survey}. Wu et al.~\cite{wu2021making} propose constructing a model of reviewer bidding behavior using the submitted bids and using this model to effectively remove outliers in the bids. They also specifically defend against groups of colluding reviewers by searching for the worst-case colluders for each reviewer and removing these potential colluders from the model's ``training data'' for that reviewer. For an overview and comparison of these techniques, see~\cite{jecmen2022tradeoffs}. 

Similar problems of malicious behavior have been studied outside of the peer review setting. 
Fraudulent reviews are a major concern on platforms like Yelp and Amazon, spurring research on methods for fraud detection in these settings~\cite{
Akoglu2013OpinionFD,Kumar2018REV2FU,Eswaran2017ZooBPBP}. These settings notably differ from ours in that reviewers are not assigned items to review: we focus on malicious behavior in the paper bidding phase, which has no analogue in product review settings. Within the crowdsourcing literature, detecting and mitigating malicious behavior by workers is the subject of some research, which often proposes using careful task design in addition to ``gold standard'' questions~\cite{Gadiraju2015UnderstandingMB,Eickhoff2012IncreasingCR}. While these techniques make low-effort responses more costly for malicious workers, the behavior of malicious reviewers in peer review is aimed specifically at manipulating the paper assignment rather than minimizing effort.

Several datasets containing conference bidding information are publicly available for research use. The PrefLib library~\cite{MaWa13a} contains a few datasets with bidding data from real AI conferences. Wu et al.~\cite{wu2021making} provide a synthetic conference dataset including synthetic bidding data, constructed by analyzing the text and citations of a large set of recent AI papers. Xu et al.~\cite{xu2018strategyproof} also reconstruct similarities for the papers and reviewers at the ICLR 2018 conference, but this dataset contains only text-similarity scores and not bidding data. 
However, these datasets crucially lack labels of which reviewers are acting maliciously, or are constructed under the assumption that all reviewers act honestly. This necessitates researchers to implement any malicious behavior synthetically (as was done in both~\cite{jecmen2020manipulation,wu2021making}). In contrast, our dataset contains data from human participants labeled with whether they were acting honestly or maliciously. A few sources~\cite{Vijaykumar_2020,littman2021collusion} discuss specific incidents of bid manipulation in real conferences, but provide only high-level details and not a structured dataset.

One of our proposed algorithms to detect malicious bidding relies on the assumption that the matrix of honest bids is approximately low-rank. Fiez et al.~\cite{fiez2020super} heuristically show that the aforementioned ICLR 2018 similarity matrix ($2435$ reviewers, $935$ papers) is approximately rank-10. Jecmen et al.~\cite{jecmen2021nearoptimal} leverage the low-rank structure of similarities to design algorithms for two-stage paper review processes.

\section{Dataset} \label{sec:data}
We first describe the data collection process, and then the contents of the collected data. The dataset is available at \url{https://github.com/sjecmen/malicious_bidding_dataset}, with full documentation given in Appendix~\ref{apdx:doc}.

\begin{table}[t!]
\centering
\begin{tabular}{lrr} \toprule
Subject area topic & \# Papers  & \# Reviewers  \\ \midrule
Humans and AI & 3 & 9 \\
Social choice theory & 3 & 11 \\
Game theory & 7 & 16 \\
Probabilistic modeling & 3 & 11 \\
Search & 3 & 7 \\
Optimization & 3 & 4 \\
Machine learning & 6 & 12 \\
\bottomrule\\
\end{tabular} 
\caption{Distribution of high-level subject area topics among the 28 papers and the 31 reviewers that completed the activity. Note that each reviewer can have up to 3 subject area topics.}\label{tab:areas} 
\end{table}

\subsection{Data Collection Process} \label{sec:coll} 
This dataset was collected as part of an voluntary, ungraded activity conducted in a graduate-level course on artificial intelligence at Carnegie Mellon University, during the game theory component of the course. Participants were students enrolled in the course, primarily PhD students in computer science. Although our data is not from a real conference, this participant population is not very different from the population of reviewers in computer science conferences: for example, 33\% of reviewers at NeurIPS 2016 were PhD students~\cite{shah2017design}.  Potential participants were explicitly informed that ``the activity is optional and won’t affect your grade in any way'' before consenting to participate. The full instructions given to participants are available in Appendix~\ref{apdx:instruct}. 

In the activity, participants act as reviewers during the paper bidding phase of a mock AI conference. 
Before the activity began, some setup was required. First, we constructed a list of 25 AI topics to use as ``subject areas'' similar to those in real conferences; these subject areas were grouped into seven high-level ``subject area topics''. Potential participants (i.e., students in the class) were then polled to ask for their areas of interest among these subject areas. 56 out of 61 total students responded to this poll. Based on these responses, we constructed a list of 28 fake paper titles; these titles were chosen so that the distribution of paper subject areas would match the distribution of participant interest. In Table~\ref{tab:areas}, we display the distribution of subject area topics for papers and for the subset of reviewers that completed the activity. 

The next step in setup was to create ``reviewer profiles'' for each potential participant. We chose three subject areas for each participant as their areas of expertise as a reviewer, chosen to match their true areas of interest as much as possible. We chose one paper for each reviewer from within one of their subject areas as the paper authored by that reviewer; papers were chosen so that each paper was the authored paper of two reviewers.

\begin{figure}[t!] 
    \centering\includegraphics[width=0.4\textwidth]{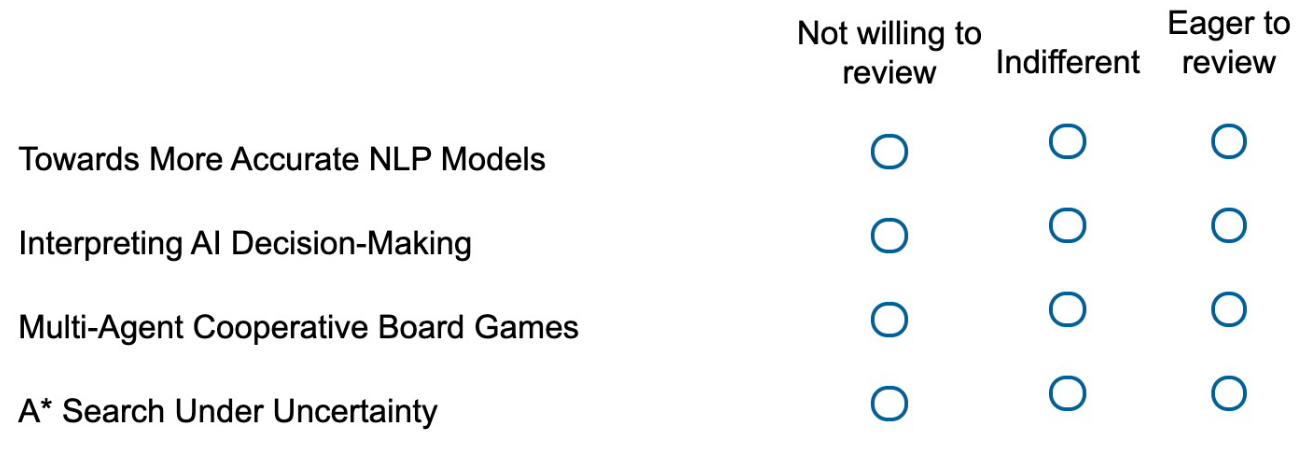}\caption{Illustration of the participant bidding interface.}\label{fig:screen}
\end{figure}

Finally, we placed reviewers into colluding groups. We made six groups of size two, two groups of size three, and eight groups of size four, leaving six participants to act solo without a colluding group. Groups were chosen so that the subject areas of the reviewers in each group overlapped as much as possible and so that reviewers with the same authored paper were not in the same group. For the six reviewers without colluders, we assigned them each a target paper from one of their subject areas; this could represent colluding with an author reviewing at a different venue or targeting a paper with the intent to ``torpedo review''~\cite{barroga2014safeguarding,paolucci2014mechanism,akst2010hate}. Each participant was provided with all information in their reviewer profile before beginning the activity, including contact information for the other participants in their colluding group.

The activity took place in two phases, so that each participant could act as an honest reviewer in the first phase and as a malicious reviewer in the second phase. 
In each phase of the activity, each participant was presented with the list of 28 paper titles and asked to submit a bid on each one. Bids were chosen from the options ``Not willing to review'', ``Indifferent'', and ``Eager to review.'' 
The bidding interface is shown in Figure~\ref{fig:screen}. After bids were placed, they could provide text responses to questions regarding their strategy. Participants were told that bids would be used to determine a paper assignment, where each paper would be assigned three reviewers and each reviewer would be assigned three papers.
\begin{itemize}[leftmargin=*]
\item In phase one, participants were instructed to bid on papers ``according to your own personal interests, as if you were actually going to review the assigned papers.'' 
\item In phase two, participants were instructed to work with their groups to manipulate the paper assignment. Specifically, participants were instructed to bid ``so that you are assigned to review each other’s papers.'' Participants were also instructed to coordinate their strategy with their groups and were free to use any method of communication to do so. Reviewers without a group were instructed to bid in order to get assigned to their given target paper. Reviewers were also told that conference program chairs were investigating the bidding for suspicious behavior: ``If they notice any reviewers bidding suspiciously, they can manually modify the assignment to their liking. For example, the PCs may look through the bids to notice any reviewers that bid positively only on a single paper and choose to ignore those bids.''
\end{itemize}
In what follows, we use the terms ``honest reviewer'' and ``malicious reviewer'' to refer to participants acting in the respective role.

\begin{figure*}[t!] 
    \centering
    \begin{subfigure}{0.48\textwidth}\includegraphics[width=1\textwidth]{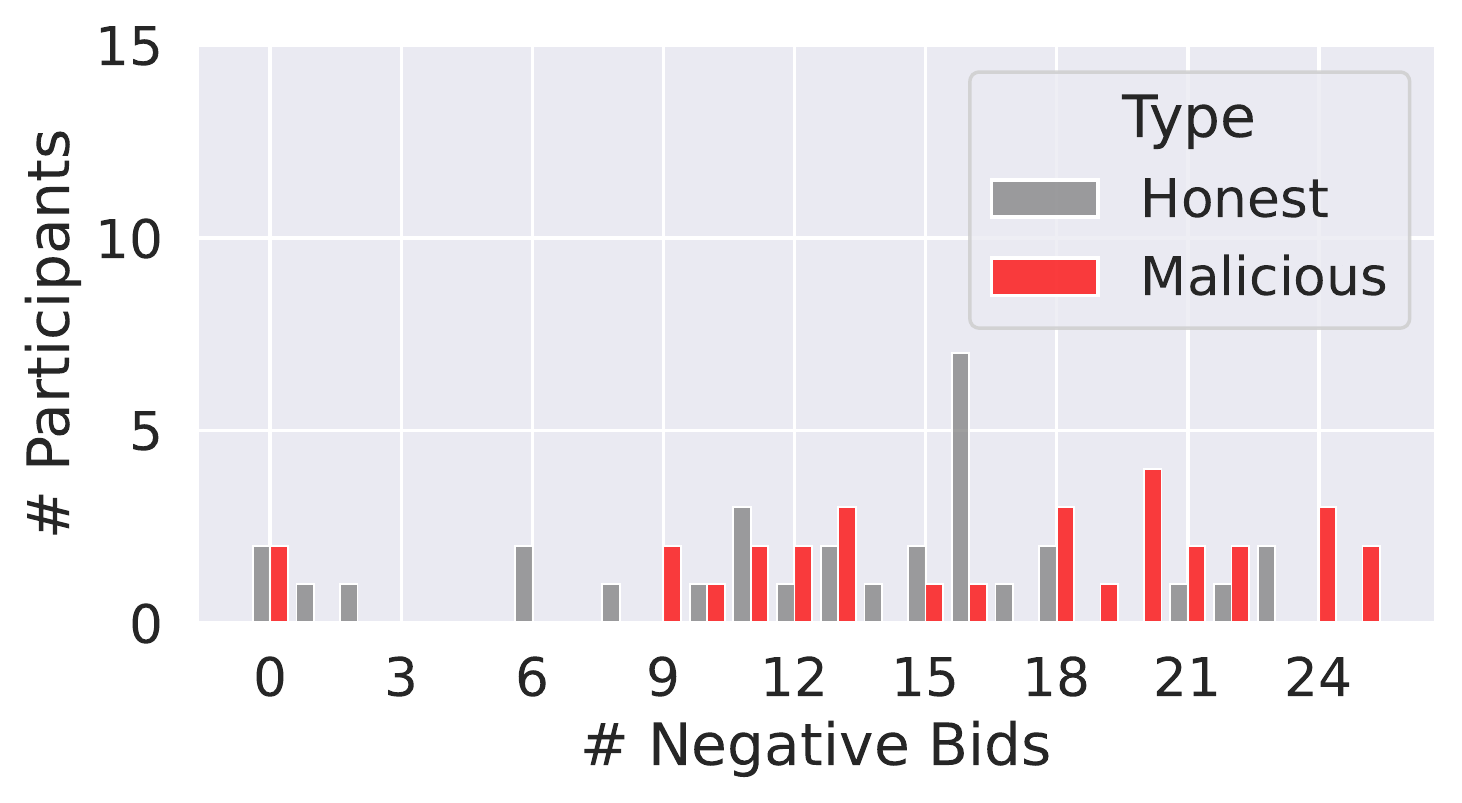}\caption{Negative bids}\label{fig:dist_neg} \end{subfigure} 
    \begin{subfigure}{0.48\textwidth}\includegraphics[width=1\textwidth]{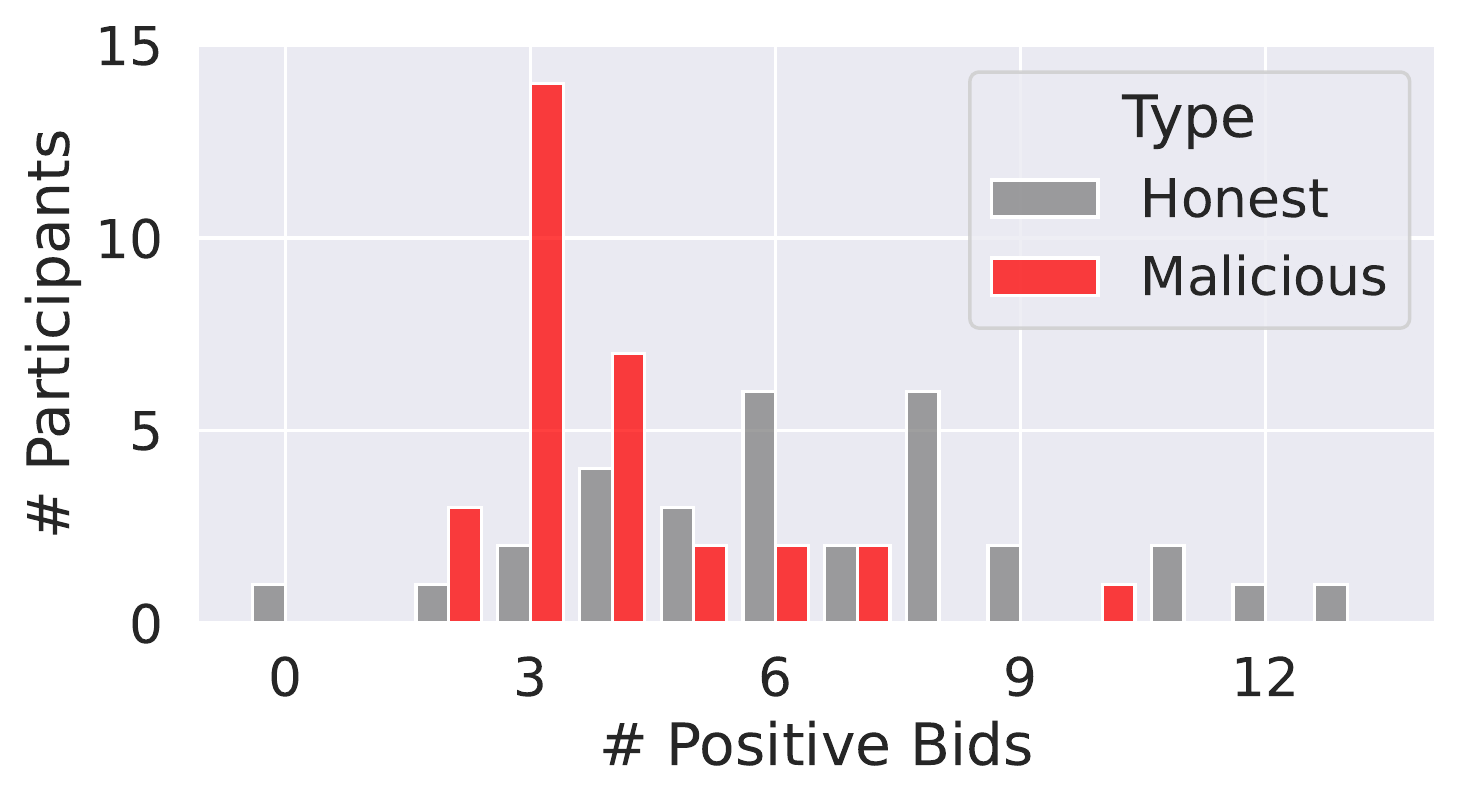}\caption{Positive bids}\label{fig:dist_pos} \end{subfigure} 
    \caption{Distributions of positive and negative bids.} \label{fig:bid_dist} 
\end{figure*}

\subsection{Dataset Contents} \label{sec:raw} 
Each of the 56 participants was given a ``reviewer profile'' as described above, consisting of three subject areas and an authored paper. Each profile also specified the participant's group, as well as a target paper if they had no colluders. We henceforth will also use the term ``target papers'' to refer to the papers authored by a reviewer's fellow group members. 
Of the 56 participants, we received 35 responses to phase one of the activity (the honest bidding) and 31 responses to phase two (the malicious bidding). In each phase, each response includes a set of 28 bids (one per paper) and a few text responses to short-answer questions asked after the bidding. Bids were either ``Not willing to review'', ``Indifferent'', or ``Eager to review''; a missing bid on a paper was interpreted as an ``Indifferent'' bid. We henceforth refer to these as ``negative'', ``neutral'', and ``positive'' bids respectively. In both phases, participants were asked: ``Did you follow any kind of strategy when bidding and if so, what was it?'' In phase two, participants were additionally asked ``Did you communicate with your other group members and if so, what did you discuss?'' and ``Do you have any thoughts on how to prevent this kind of malicious behavior in conferences?'' 
The dataset was de-identified by course instructors.

\section{Description of Bidding Behavior} \label{sec:desc}
In this section, we provide descriptions of the collected bidding data. We first quantitatively consider the bidding data itself, and then qualitatively analyze the strategy descriptions provided by participants. 

\subsection{Quantitative Description}

In Figure~\ref{fig:bid_dist}, we compare the distributions of positive and negative bids between the honest reviewers and the malicious reviewers. Each bar corresponds to a specific number of positive or negative bids, with the height of the bar indicating the number of honest or malicious reviewers submitting that many bids. In Figure~\ref{fig:dist_neg}, we see that malicious reviewers generally provided more negative bids than honest reviewers, although both honest and malicious reviewers provided high numbers of negative bids. We see in Figure~\ref{fig:dist_pos} that honest reviewers generally gave more positive bids than malicious reviewers. Nearly half of the malicious reviewers (14 responses) bid positively on exactly three papers (the number of papers they will be assigned), whereas honest reviewers were much more likely to bid on additional papers.

\subsection{Qualitative Description} \label{sec:strats} 
During both phases of the activity, participants were asked to describe any strategy they used in a text response. We analyze these responses in conjunction with the actual provided bids in order to determine what strategy participants implemented. These strategies can be used by researchers to realistically ``scale up'' the dataset, as we demonstrate in Section~\ref{sec:synth}.

When providing honest bids, almost everyone (32 responses) specified their strategy as some form of ``bidding based on my interests.'' A few of the comments were more detailed about an additional strategy they followed to get assigned high-expertise papers (e.g., bid positively on exactly three papers), but none of these were common.

For the malicious bids, we categorize them into five broad ``strategies.'' Below we describe the prototypical aspects of each strategy, although participants' actual implementations varied. 
\begin{enumerate}[leftmargin=*] 
    \item \textit{Basic:} On target papers, bid mostly positively and sometimes neutral. On non-target papers within the reviewer's subject area, bid mostly neutral with a few positive bids. (11 people)
    \item \textit{Negative-in-area:} As in \textit{Basic}, but bid mostly negatively on non-target papers within the reviewer's subject area (still with a few positive bids). (9 people)
    \item \textit{Overlap:} As in \textit{Basic}, but coordinate the bids on non-target papers with other group members. Specifically, all group members bid positively on the same set of non-target, in-area papers (with the aim of ``overloading'' them so that not all group members can be assigned to them). (3 people)
    \item \textit{Cycle:} As in \textit{Basic}, but bid in a particular manner on target papers. Specifically, each group member bids positively on one other member's target paper to create a cycle of positive bids. Group members bid neutral or negative on other target papers. (4 people)
    \item \textit{Popularity:} As in \textit{Basic}, but choose bids on non-target, in-area papers based on perceived paper ``popularity''. Specifically, bid positively on a small number of non-target papers that are likely to be unpopular among other reviewers (with the aim of being assigned to these in addition to the target papers). Bid neutral on non-target papers that are likely to be popular among other reviewers (with the aim of not being assigned to them). (2 people)
\end{enumerate} 
Two other participants did not describe or implement an understandable strategy. Additionally, 19 of these responses specified that they coordinated with their group in forming their strategy. 

Our choice of strategy categorization is not unique, as participant strategies could further be broken down on the basis of additional information. Some strategies specified how they chose the number of positive bids, usually in consideration of the fact that each reviewer would be assigned three papers. Some strategies specified how they chose which non-target papers to bid positively on (e.g., only those outside their subject area). Some strategies bid positively on all target papers while others split between positive and neutral. 
We choose to focus on the above categorization, leaving analysis on the basis of these additional factors for future work.

\section{Evaluation of Bidding Behavior} \label{sec:exps}
In this section, we analyze the performance of malicious reviewers empirically in terms of successfully manipulating the assignment and avoiding detection. We also consider the performance of several baseline detection algorithms. 
Specifically, we run two empirical evaluations: one which examines how successful each reviewer is at manipulating the assignment, and one which examines how well each reviewer avoids detection by simple detection algorithms. 
We run multiple trials of each evaluation, where each trial considers one group of malicious reviewers. 
In each trial, we construct a full set of reviewers by taking the malicious reviewers in the group under consideration and adding honest reviewers at random until the number of reviewers equals the number of papers ($28$). We then use this set of reviewers along with the fixed set of $28$ papers for the experiment. In total, we run $100$ trials of each evaluation for each group, aggregating results over the $100$ trials.

\subsection{Manipulation Success Evaluation} \label{sec:successeval}
Malicious reviewers were instructed to bid in order to get assigned to their target papers, but were not given a specific numerical objective to optimize. Therefore, some participants may have divided the target papers among their group while others simultaneously targeted all target papers. In our analysis, we define a ``successful manipulation'' as when a malicious reviewer is assigned to at least one of their target papers. This definition is reasonable for different forms of strategic coordination within groups and is robust to the non-participation of a reviewer's group members.

We compute similarities $\simmat_{\adrev, \adpap} \in \mathbb{R}$ between each reviewer $\adrev$ and paper $\adpap$ as follows. We first compute a subject area score $\areamat_{\adrev, \adpap}$ by comparing the reviewer subject areas and paper subject areas. If the paper's subject area matches one of the reviewer's subject areas, we set $\areamat_{\adrev, \adpap} = 1$. Otherwise, if the paper's high-level subject area topic matches one of the reviewer's, we set $\areamat_{\adrev, \adpap} = 0.5$. Else, we set $\areamat_{\adrev, \adpap} = 0$. Bid values $\bidmat_{\adrev, \adpap} \in \{-1, 0, 1\}$ are set corresponding to negative, neutral, and positive bids respectively. Final similarities are then computed as $\simmat_{\adrev, \adpap} = (1 + \areamat_{\adrev, \adpap}) 2^{\bidmat_{\adrev, \adpap}}$, modeled off of the similarity formula used for NeurIPS 2016~\cite{shah2017design}.

After computing similarities, we then find a maximum-similarity assignment of reviewers to papers~\cite{charlin13tpms, charlin2012framework, goldsmith07aiconf, flach2010kdd, kobren19localfairness}, subject to constraints that each paper is assigned three reviewers, each reviewer is assigned to three papers, and no reviewer is assigned to a paper they authored. This method of assigning reviewers to papers is standard in modern conferences~\cite[Section 3]{shah2021survey}. In the resulting assignment, we determine whether each malicious reviewer was successfully assigned to at least one of their target papers.

\begin{figure*}[t!] 
    \centering
    \begin{subfigure}{0.45\textwidth}\includegraphics[width=1\textwidth]{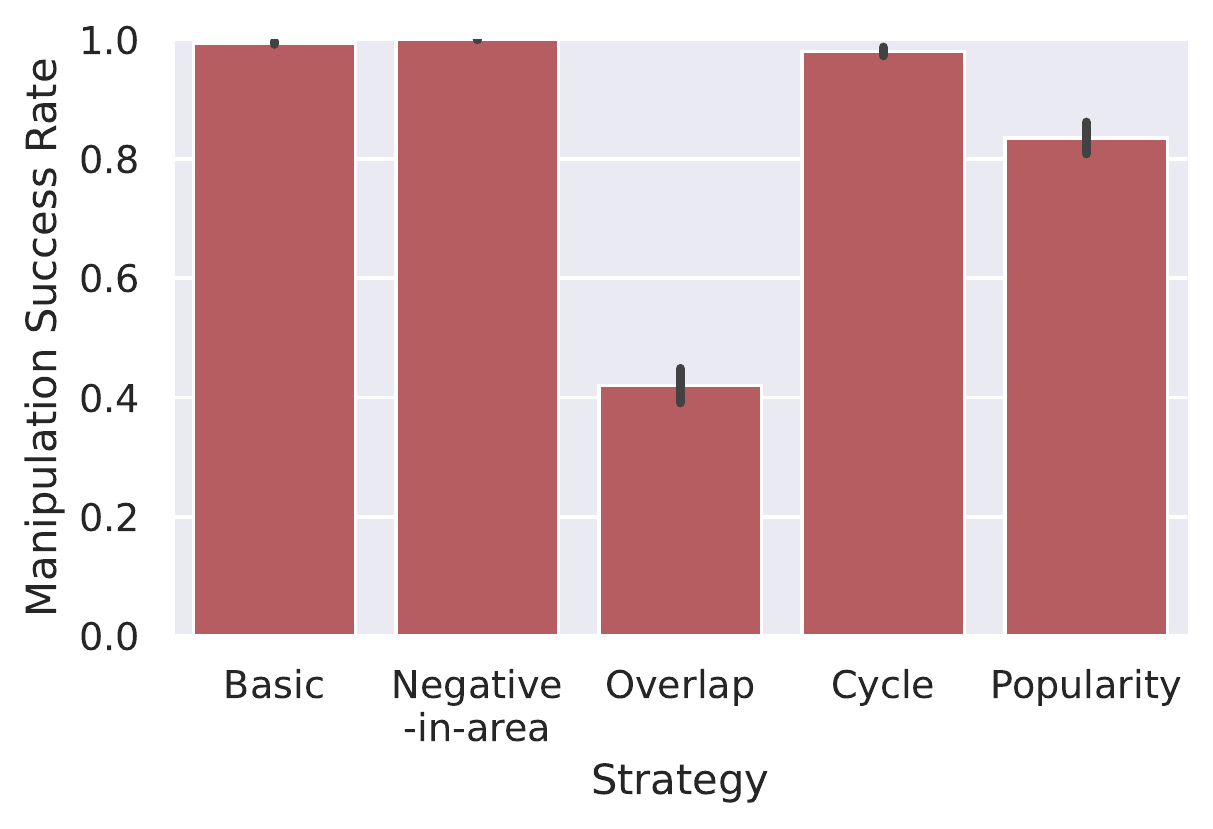}\caption{Malicious reviewer success rate}\label{fig:success}
    \end{subfigure} \quad
    \begin{subfigure}{0.45\textwidth}\includegraphics[width=1\textwidth]{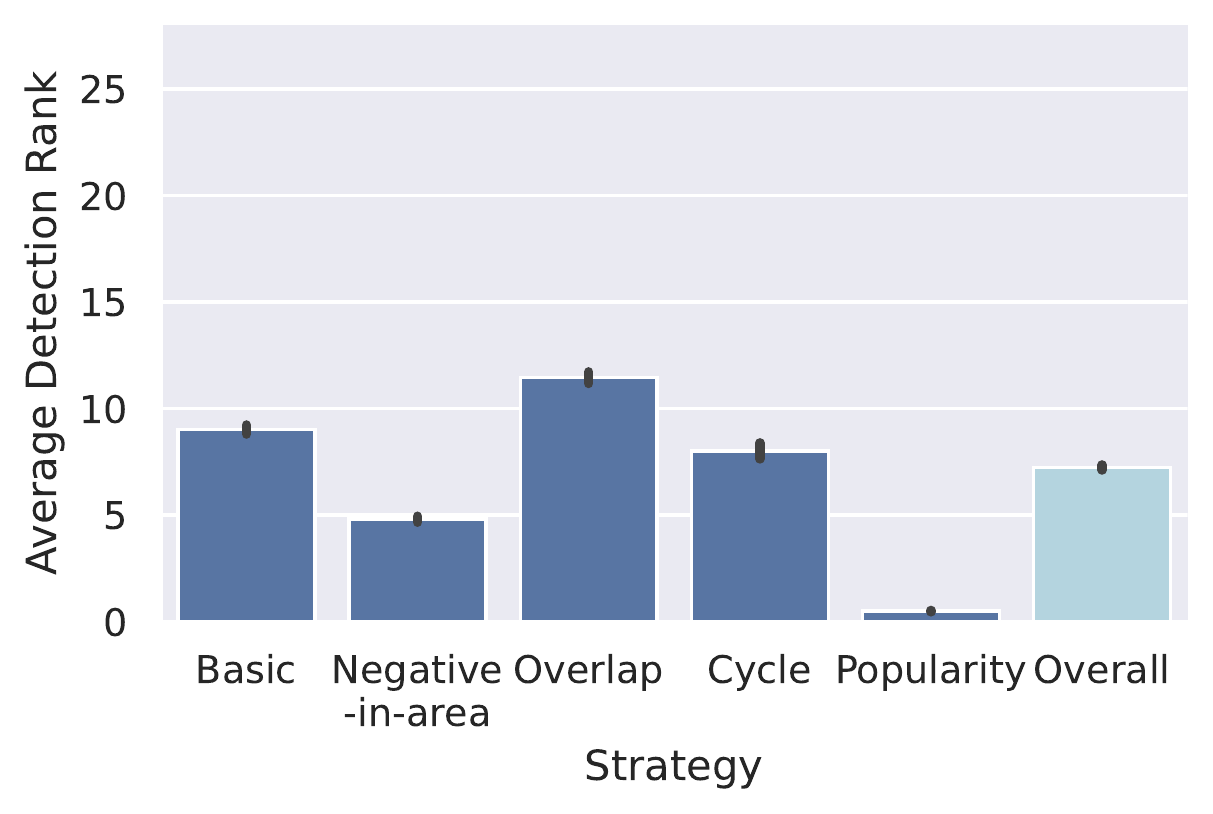}\caption{Counting detection}\label{fig:rank_counting} \end{subfigure} \\
    \begin{subfigure}{0.45\textwidth}\includegraphics[width=1\textwidth]{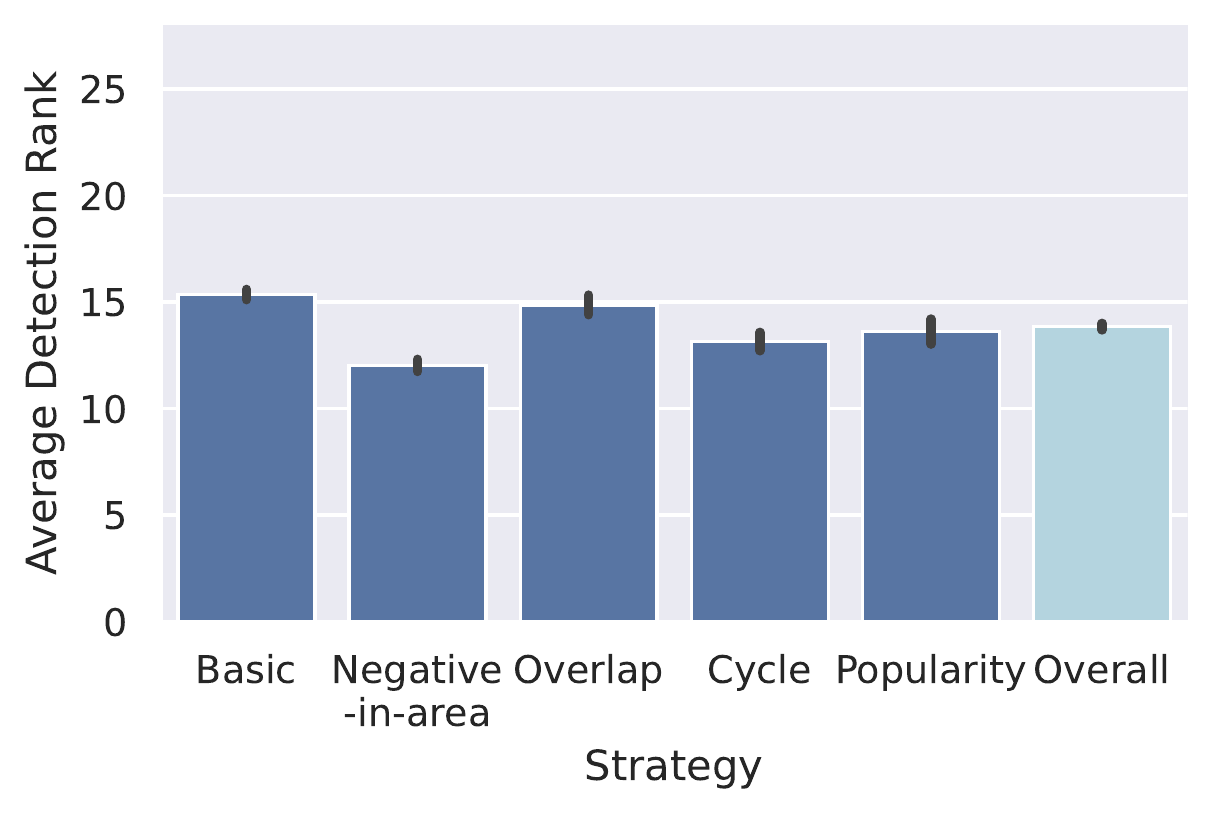}\caption{Ring detection}\label{fig:rank_cluster} \end{subfigure} \quad
    \begin{subfigure}{0.45\textwidth}\includegraphics[width=1\textwidth]{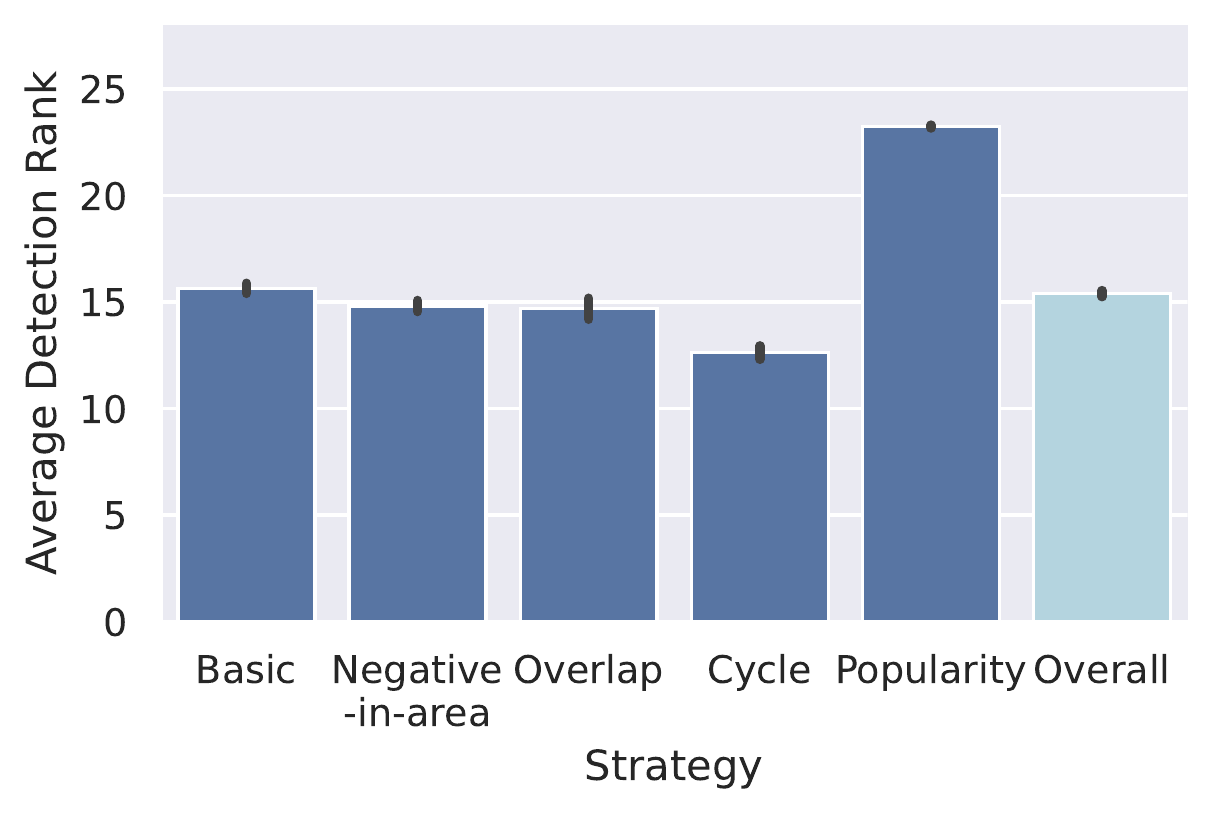}\caption{Low-rank detection}\label{fig:rank_lowrank} \end{subfigure} 
    \caption{Average success rate of manipulation strategies (Figure~\ref{fig:success}) and average rank of malicious reviewers under different detection algorithms (Figures~\ref{fig:rank_counting}-\ref{fig:rank_lowrank}). In Figures~\ref{fig:rank_counting}-\ref{fig:rank_lowrank}, 
    a lower rank value indicates that the algorithm has marked the malicious reviewer as more suspicious.} \label{fig:actual} 
\end{figure*}

In Figure~\ref{fig:success}, we display the results from this empirical evaluation. Each bar represents the average success rate of all reviewers with a given strategy, with error bars representing standard error of the mean. The two strategies used by the greatest number of reviewers (\textit{Basic} and \textit{Negative-in-area}) had a perfect success rate, indicating that the additional sophistication of the \textit{Negative-in-area} strategy was not necessary to secure the target assignments. 
The \textit{Cycle} strategy was also highly successful, likely due to its similarity to \textit{Basic}. 
The \textit{Overlap} strategy had a very low success rate despite also being very similar to the \textit{Basic} strategy; we suspect this was due to poor implementation of the strategy, since only a few participants attempted this strategy. The \textit{Popularity} strategy had mixed success, perhaps due to the difficulty of correctly guessing which papers are popular. 

\subsection{Detection Evaluation} \label{sec:detecteval}
We consider three simple detection algorithms as baselines for analysis. Each algorithm outputs a ranking of all reviewers, where reviewers higher in the ranking are deemed more likely to be malicious. Before running any algorithm, bids made by each reviewer on their own paper were discarded. 
\begin{itemize}[leftmargin=*]
    \item \textit{Counting Detection}: Compute a score for each reviewer equal to the number of negative bids minus the number of positive bids. Output the ranking of all reviewers by score (where higher scores are higher-ranked).
    \item \textit{(Pairwise) Ring Detection}:  
    We first construct a ranking of all pairs of reviewers. Pairs in which both reviewers bid positively on each others' papers are ranked the highest, followed by pairs in which exactly one reviewer bid positively on the other's paper, followed by pairs in which neither reviewer bid positively on the other's paper. Within each of these categories, we rank the pairs of reviewers as follows. For each reviewer in the pair, count the number of negative bids made and subtract the number of positive bids, not counting the bid made on the paired reviewer's paper. Rank the pairs within each category by the total score for the two reviewers in the pair. Finally, output a ranking of reviewers by their highest position in the ranking of pairs (breaking ties arbitrarily).
    \item \textit{Low-Rank Detection}: 
    This algorithm leverages the assumption that the bidding matrix $\bidmat$ (with entries $\bidmat_{\adrev, \adpap}$ defined in Section~\ref{sec:successeval}) is approximately low-rank; this assumption is discussed in Section~\ref{sec:relwork}.
    Compute a low-rank approximation $L$ to $\bidmat$.  
    Compute a score for each reviewer $\adrev$ as $\sum_{\adpap \in \text{Papers}} | \bidmat_{\adrev, \adpap} - L_{\adrev, \adpap}|$ (where the sum is taken over all papers). 
    Output the ranking of all reviewers by this score. 
    Intuitively, this method classifies bids as anomalous when they cannot be represented well in a low-rank matrix.  
    In our experiments, we use a rank-3 approximation; this number was chosen heuristically based on inspection of the singular values of several sampled honest bidding matrices.
\end{itemize}
In each trial of this evaluation, we select a set of reviewers as described at the beginning of Section~\ref{sec:exps} and collect the bids of each selected honest or malicious reviewer. We then run a detection algorithm on these bids and report the output rank of each malicious reviewer.

In Figures~\ref{fig:rank_counting}-\ref{fig:rank_lowrank}, we display the average ranks of malicious reviewers output by the detection algorithms, with error bars representing standard error of the mean. Rank values (on the y-axis) indicate the number of other reviewers marked as more suspicious than the malicious reviewer, ranging between $0$ (most suspicious) and $27$ (least suspicious); a lower rank value indicates that the algorithm was more successful at detecting the manipulation. 
In each plot, the five leftmost bars display the average ranks of malicious reviewers using each strategy under a given detection algorithm. 
Between the two most popular strategies (\textit{Basic} and \textit{Negative-in-area}), all three detection algorithms were better at detecting the \textit{Negative-in-area}. \textit{Negative-in-area} was particularly well-detected by the \textit{Counting Detection} algorithm, since it specifically targets the kind of behavior done by the \textit{Negative-in-area} strategy. \textit{Counting Detection} also does very well against the \textit{Popularity} strategy, although this is simply because both participants implementing this strategy happened to bid negatively on nearly all papers outside of their subject area. 
The rightmost bar in each plot shows the overall performance of the detection algorithms by averaging the output ranks across all malicious reviewers. \textit{Counting Detection} has decent performance overall, averaging ranking malicious reviewers around the $25$th percentile. The other algorithms do poorly, averaging ranking malicious reviewers around the $50$th percentile (essentially no better than random). Some malicious reviewers consciously avoided seeming ``too connected'' to their group by bidding neutral on some target papers, hurting the performance of \textit{Ring Detection}; this algorithm may also have suffered from the small size of the dataset, since pairs of honest reviewers are likely to bid positively on each others' papers by chance.

\begin{figure*}[t!] 
    \centering
    \begin{subfigure}{0.48\textwidth}\includegraphics[width=1\textwidth]{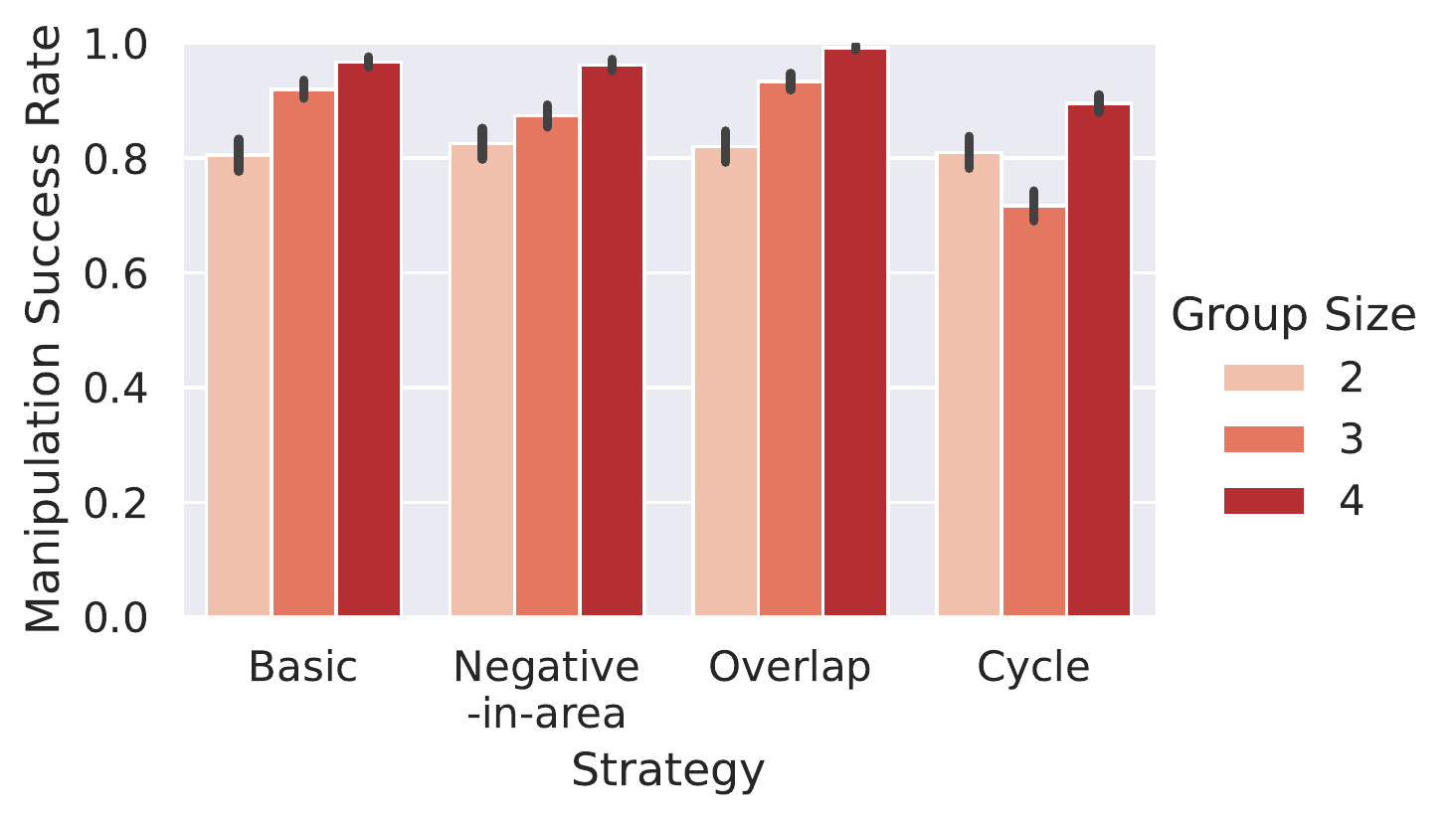}\caption{Malicious reviewer success rate}\label{fig:success_synth} \end{subfigure} 
    \begin{subfigure}{0.48\textwidth}\includegraphics[width=1\textwidth]{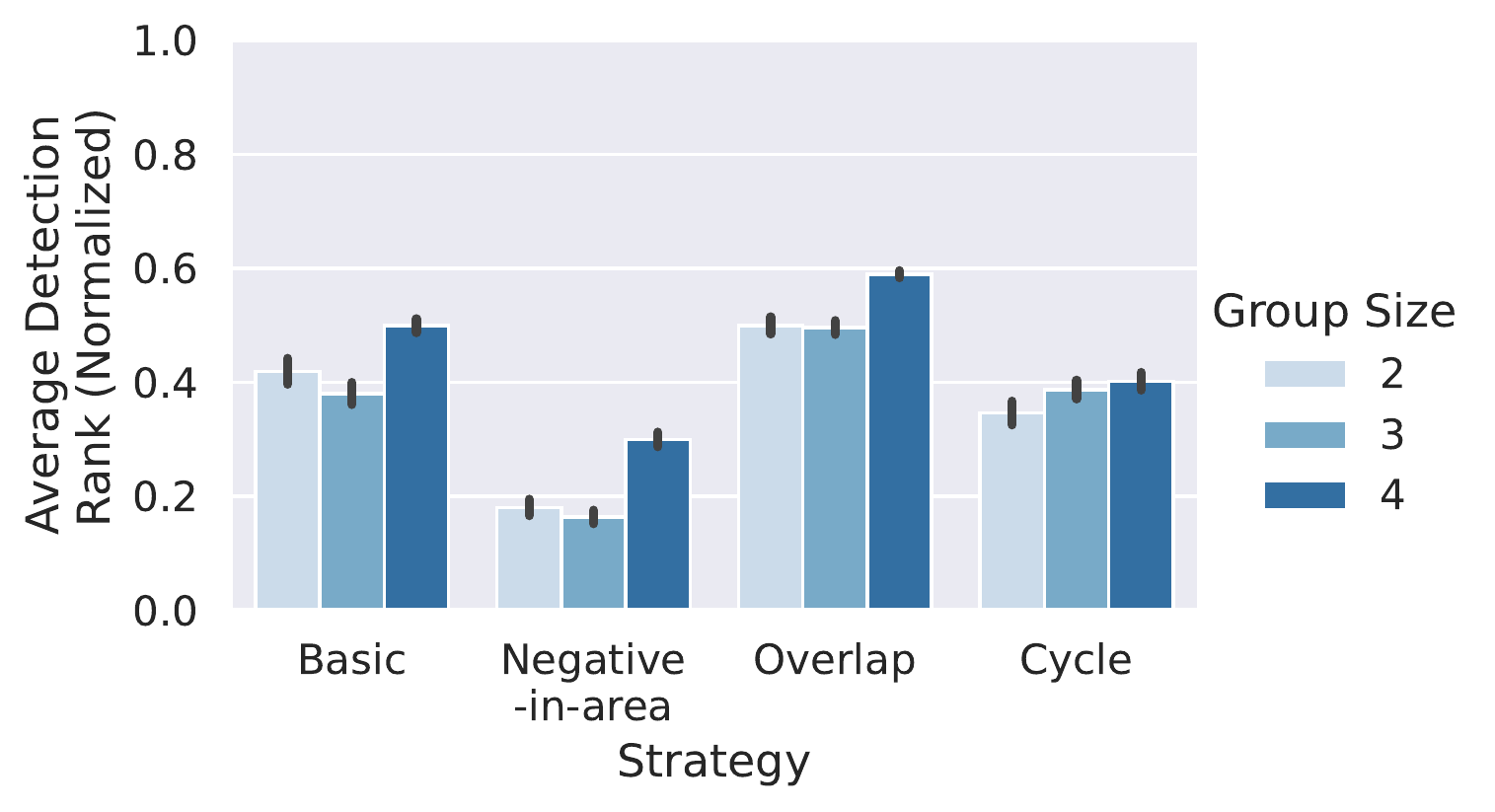}\caption{Counting detection}\label{fig:rank_counting_synth} \end{subfigure} \\
    \begin{subfigure}{0.48\textwidth}\includegraphics[width=1\textwidth]{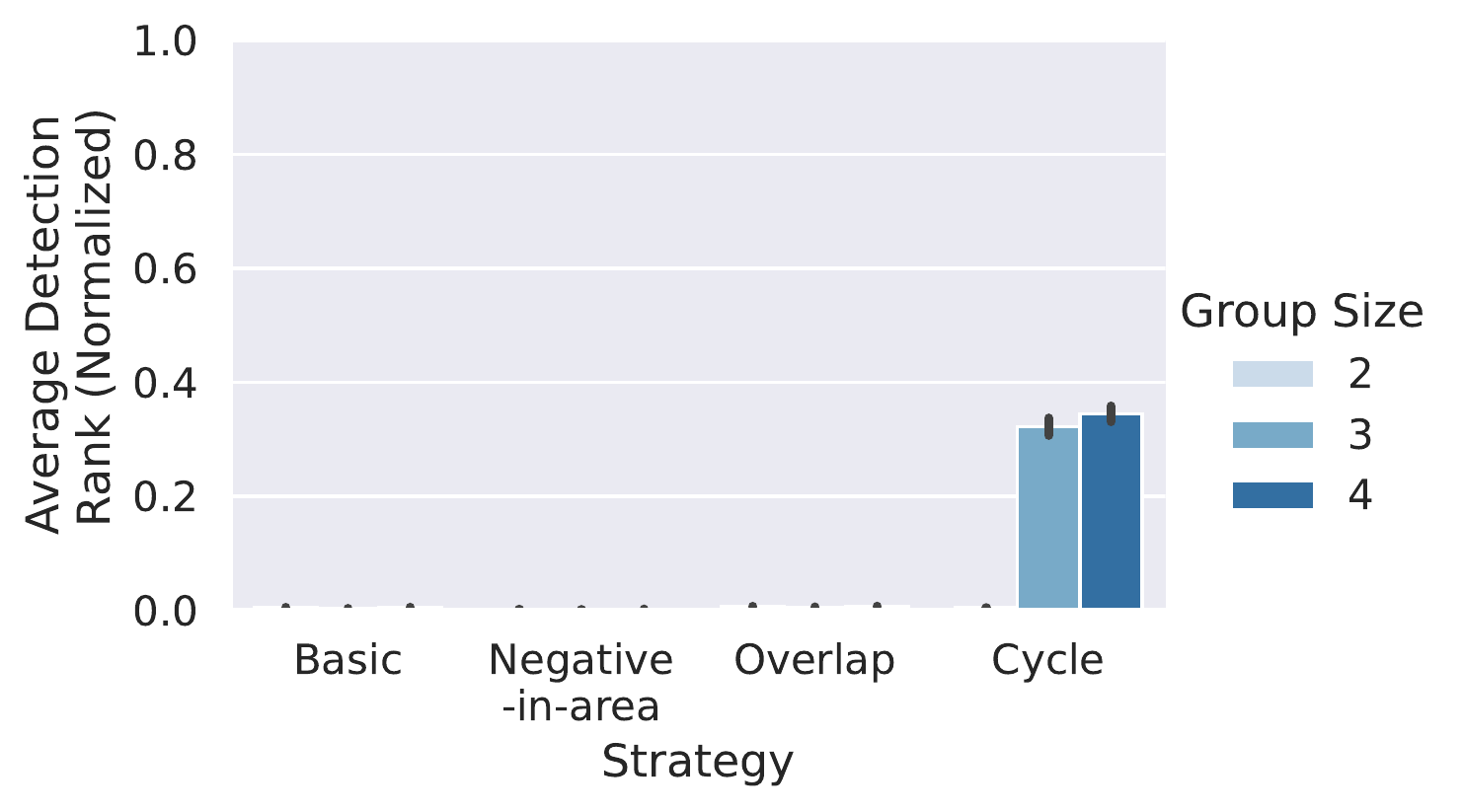}\caption{Ring detection}\label{fig:rank_cluster_synth} \end{subfigure} 
    \begin{subfigure}{0.48\textwidth}\includegraphics[width=1\textwidth]{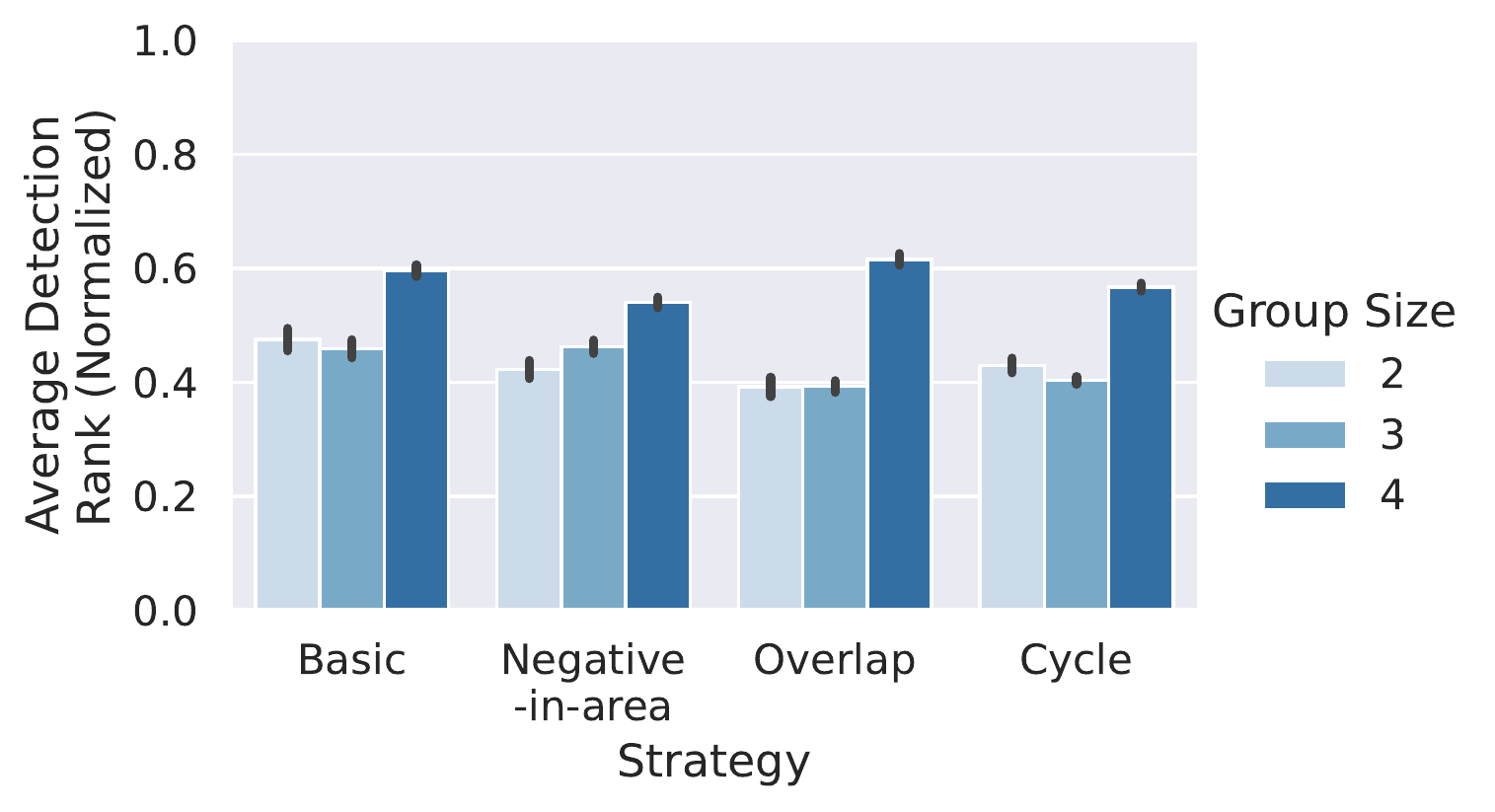}\caption{Low-rank detection}\label{fig:rank_lowrank_synth} \end{subfigure} 
    \caption{Results from synthetic scaled-up experiments with 5000 reviewers and papers. Different colors indicate different malicious group sizes. 
    In Figures~\ref{fig:rank_counting_synth}-\ref{fig:rank_lowrank_synth}, a lower normalized rank indicates the malicious reviewer was detected as more suspicious.  } 
    \label{fig:synth} 
\end{figure*}

\section{Analysis of Synthetically Scaled-up Data}\label{sec:synth}
In this section, we run experiments on synthetically scaled-up versions of the data. We first describe the procedure we use to scale up the data, followed by the experimental results. 

\subsection{Synthetic Dataset Construction} 
We construct a large-scale synthetic dataset based on characteristics of our collected data. 
Using the strategy categorization introduced in Section~\ref{sec:strats}, we wrote our own implementation of each strategy (other than \textit{Popularity}), modeling any bidding behavior not specified by the strategy after a random reviewer in the original dataset. We remark that this procedure is only one example of how our dataset can be used to inform the creation of larger-scale synthetic datasets and is far from the only way to do so. 

We first choose the number of reviewers and papers in the scaled-up data, the malicious group size (2, 3, or 4), and the strategy that the malicious reviewers will employ. For simplicity, authorship will be one-to-one between papers and reviewers, so each reviewer corresponds to one authored paper. 

We then construct the reviewer and paper subject areas. To determine the subject areas of the malicious reviewers and the papers authored by them, we randomly choose a malicious group of the chosen size from the original data. We then copy the subject areas of these reviewers and their authored papers. Similarly, to determine subject areas for each honest reviewer and their authored paper, we randomly choose any reviewer from the original data and copy the subject areas of the reviewer and their authored paper. 

We next construct the bids for each honest reviewer. For each honest reviewer, we randomly choose an honest reviewer in the original data to use as a ``model''. We count the number of positive bids made by this original reviewer on papers within their high-level subject area topics, and add this many positive bids for the new reviewer on random papers within their subject area topics. We do the same for positive bids on papers outside their subject area topics, counting the number of bids made by the original reviewer and adding this many at random for the new reviewer. We then repeat this process for negative bids, but scale up the number of negative bids added by the ratio between the new and old numbers of papers. For example, if the original reviewer made two negative bids on papers within their subject area topic and we are scaling up from 28 papers to 280 papers, we would add 20 negative bids on papers within the new reviewer's subject area topics. We choose to scale up the number of negative bids and not the number of positive bids because the reviewer loads are still three in the new experiments despite the larger number of papers. Many reviewers considered the reviewer loads in choosing how many positive bids to place (e.g., by bidding positively on exactly three papers), and this procedure preserves that behavior. In contrast, many reviewers bid negatively on a large number of papers, suggesting that they would place even more negative bids if the number of papers was increased.

Finally, we construct bids for the malicious reviewers. For each malicious reviewer, we first construct bids on all non-target papers using the method described in the previous paragraph for honest reviewers. However, rather than randomly choosing a model reviewer from among the honest reviewers, we randomly choose a model reviewer from among the malicious reviewers with the same strategy.  We then modify the bids in a different way depending on the strategy chosen. For the \textit{Basic} strategy, we simply add a positive bid on each target paper. For the \textit{Overlap} strategy, we also add a positive bid on each target paper; we then further adjust the bids of all malicious reviewers so that the positive bids are on the same set of papers. For the \textit{Cycle} strategy, we have each reviewer bid positively on only one target paper and neutral on the others, constructing a cycle. We do not implement the \textit{Popularity} strategy due to its rarity and the difficulty of modeling how a reviewer might predict which papers are popular.

\subsection{Synthetic Results}
We run the experiments described in Section~\ref{sec:exps} on the scaled-up bids and subject areas, aggregating results over $100$ trials of synthetic dataset construction for each setting. 

In Figure~\ref{fig:synth}, we display results from both experiments for each malicious reviewer strategy when the data is scaled up to $5000$ papers and reviewers, varying the malicious group size. In Figure~\ref{fig:success_synth}, we see that all four implemented strategies are very successful at all group sizes. The \textit{Cycle} strategy is slightly less successful than the others; this is perhaps because by bidding positively on only one target paper, the strategy is less robust to the many honest reviewers also bidding on the target papers. All strategies are most successful for the largest malicious group size, likely because these groups have more target papers that can be assigned. 

In Figures~\ref{fig:rank_counting_synth}-\ref{fig:rank_lowrank_synth}, we display the detection ranks output by the three detection algorithms. The rank values (on the y-axis) are normalized by the number of reviewers so that they range from $0$ (most suspicious) to $1$ (least suspicious). 
In Figure~\ref{fig:rank_counting_synth}, we see that the \textit{Counting Detection} algorithm does moderately well at detecting the \textit{Negative-in-area} and \textit{Cycle} strategies, although it does relatively worse on all strategies as compared to the original data. This may be because both malicious and honest reviewers make relatively more negative bids than positive bids in the scaled-up data, and so the algorithm must look for a weaker signal in a larger set of reviewers than in the original data. In contrast, Figure~\ref{fig:rank_cluster_synth} shows that the \textit{Ring Detection} algorithm does extremely well at detecting malicious behavior. This is because our implementations of the \textit{Basic}, \textit{Negative-in-area}, and \textit{Overlap} strategies bid positively on all target papers, and so the detection of these clusters cuts through the noise of the many honest reviewers. \textit{Cycle} avoids detection by this algorithm when the malicious group size is greater than 2, since these malicious reviewers avoid forming pairwise rings of positive bids. Figure~\ref{fig:rank_lowrank_synth} shows that the \textit{Low-Rank Detection} algorithm performs poorly, as in the original data.

We also run additional experiments varying the total number of reviewers and papers, which we present in Appendix~\ref{apdx:addl_results}.

\section{Discussion} \label{sec:disc}
In this work, we construct and release a dataset on malicious paper bidding, along with our analysis of the behavior employed by participants. We also evaluate the effectiveness of various participant strategies and detection algorithms. 
Our dataset has been de-identified, and furthermore the risk to participants in the event of any re-identification is low since the dataset includes no sensitive information. 

One major limitation of our work is that our dataset is from a mock conference setting and may not be perfectly representative of real-world behavior. Thus, in future work, our dataset should be used as just one method of evaluation alongside others. Any proposed detection algorithm should at least be effective against the strategies identified here, but good performance on our dataset alone is not sufficient to show an algorithm’s effectiveness in practice. 
Another possible limitation of our work is that malicious reviewers could use the data we provide and any future research on it to improve their strategies. Those researching defenses against malicious behavior should consider that an adversary can adapt in response to a new defense and develop methods that are robust to adversarial changes in behavior. 

Our dataset may be useful within various directions of future work that aim to address malicious behavior in peer review. First, our work considers three algorithms for detecting malicious bidding, which we intentionally choose as very simple baselines. The vast literature on anomaly detection proposes many more complex techniques that could be adapted for our setting, including in particular algorithms for detecting fraudulent product reviews~\cite{
Akoglu2013OpinionFD,Kumar2018REV2FU,Eswaran2017ZooBPBP}. In addition to new techniques for detecting malicious bids, new algorithms for paper assignment that are robust to the presence of malicious bids (e.g., \cite{jecmen2020manipulation,wu2021making}) can be developed and evaluated using our dataset. Finally, as more techniques to address malicious behavior are proposed and deployed~\cite{jecmen2022tradeoffs}, a valuable goal for future work is to provide guidance to conference program chairs about which techniques they should deploy at their venue. For example, as one approach, the data and strategies we present could be analyzed in a game-theoretic framework to identify the optimal defensive strategy for program chairs to deploy against an adversarial group of malicious reviewers. 

The problem of bid manipulation in conferences is of great practical importance to the research community. We hope that by releasing this dataset, other researchers will more easily be able to conduct research on this issue. 

\ifisarxiv 
\section*{Acknowledgments}
\else
\begin{acks}
\fi 
This work was conducted under the approval of the Carnegie Mellon University IRB. This work was supported in part by NSF CIF 1942124 and IIS 2046640. We thank Hanrui Zhang for helpful comments and discussion.
\ifisarxiv \else
\end{acks}
\fi

\fullver{\bibliographystyle{unsrt}}
\confver{\bibliographystyle{ACM-Reference-Format}}
\bibliography{bibtex}

\newpage
\appendix

\ifisarxiv
\pagebreak
\noindent{\LARGE \bf Appendices}
\else
\fi
\section{Dataset Documentation} \label{apdx:doc}
The dataset and our analysis code can be found at \url{https://github.com/sjecmen/malicious_bidding_dataset}. 
The authors will maintain the dataset in the linked Github repository. 
This dataset is licensed under a CC BY 4.0 license.\footnote{\url{https://creativecommons.org/licenses/by/4.0/}}. 
This work was conducted under the approval of the Carnegie Mellon University IRB. 
This dataset is intended for use by other researchers, specifically on the topic of addressing malicious behavior in peer review. See Section~\ref{sec:coll} for a detailed description of the data collection process. 

The dataset consists of 2 text files and 4 CSV files. The two text files respectively list the subject areas and paper titles used in the activity. Below, we describe the format of the CSV files.
\begin{itemize}[leftmargin=*]
    \item The file `setup.csv` contains the reviewer profile information, in the following columns:
    \begin{itemize}
        \item \textbf{name}: Anonymized string ID for each potential participant.
        \item \textbf{sas}: Three space-separated integers, indicating the indices of the subject areas for this reviewer.
        \item \textbf{authored\_sa}: Subject area index of the paper authored by this reviewer.
        \item \textbf{authored\_id}: Paper title index of the paper authored by this reviewer.
        \item \textbf{target\_sa}: Subject area index of the target paper for this reviewer (if no colluders).
        \item \textbf{target\_id}: Paper title index of the target paper for this reviewer (if no colluders).
        \item \textbf{group}: Integer ID for the reviewer's group of colluders. 
    \end{itemize}
    \item The file `honest\_bidding.csv` contains the responses to the first phase of the activity on honest bidding. The \textbf{Name} column contains the participant ID for each response. The remaining columns indicate responses to the questions stated in the second header row. 
    \item The file `malicious\_bidding.csv` contains the responses to the second phase of the activity on malicious bidding, formatted in the same way.
    \item The file `strategy\_annotations.csv` contains our categorization of participant responses by strategy. The \textbf{Name} column contains the participant ID. The \textbf{Strategy} column contains an integer indicating the strategy, as an index into the strategy list [Basic, Negative-in-area, Overlap, Cycle, Popularity]; an entry of $-1$ indicates no strategy could be discerned. The \textbf{Discussed} column contains an entry in \{Y, N\} indicating whether the participant discussed their strategy with their colluders, with an empty entry indicating an unclear response.
\end{itemize}

\confver{
\begin{figure*}[t!] 
    \centering
    \begin{subfigure}{0.48\textwidth}\includegraphics[width=1\textwidth]{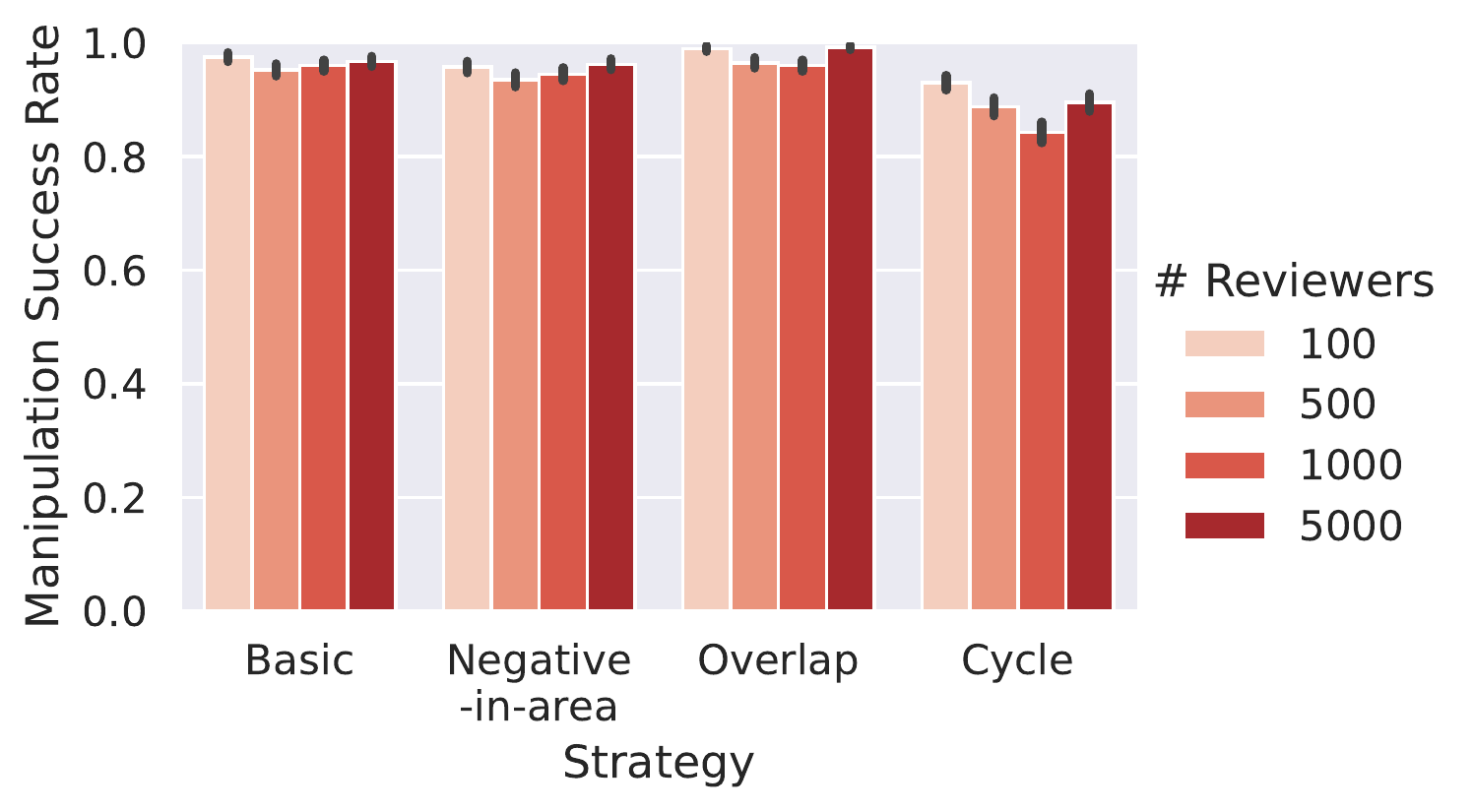}\caption{Malicious reviewer success rate}\label{fig:success_synth_n} \end{subfigure} 
    \begin{subfigure}{0.48\textwidth}\includegraphics[width=1\textwidth]{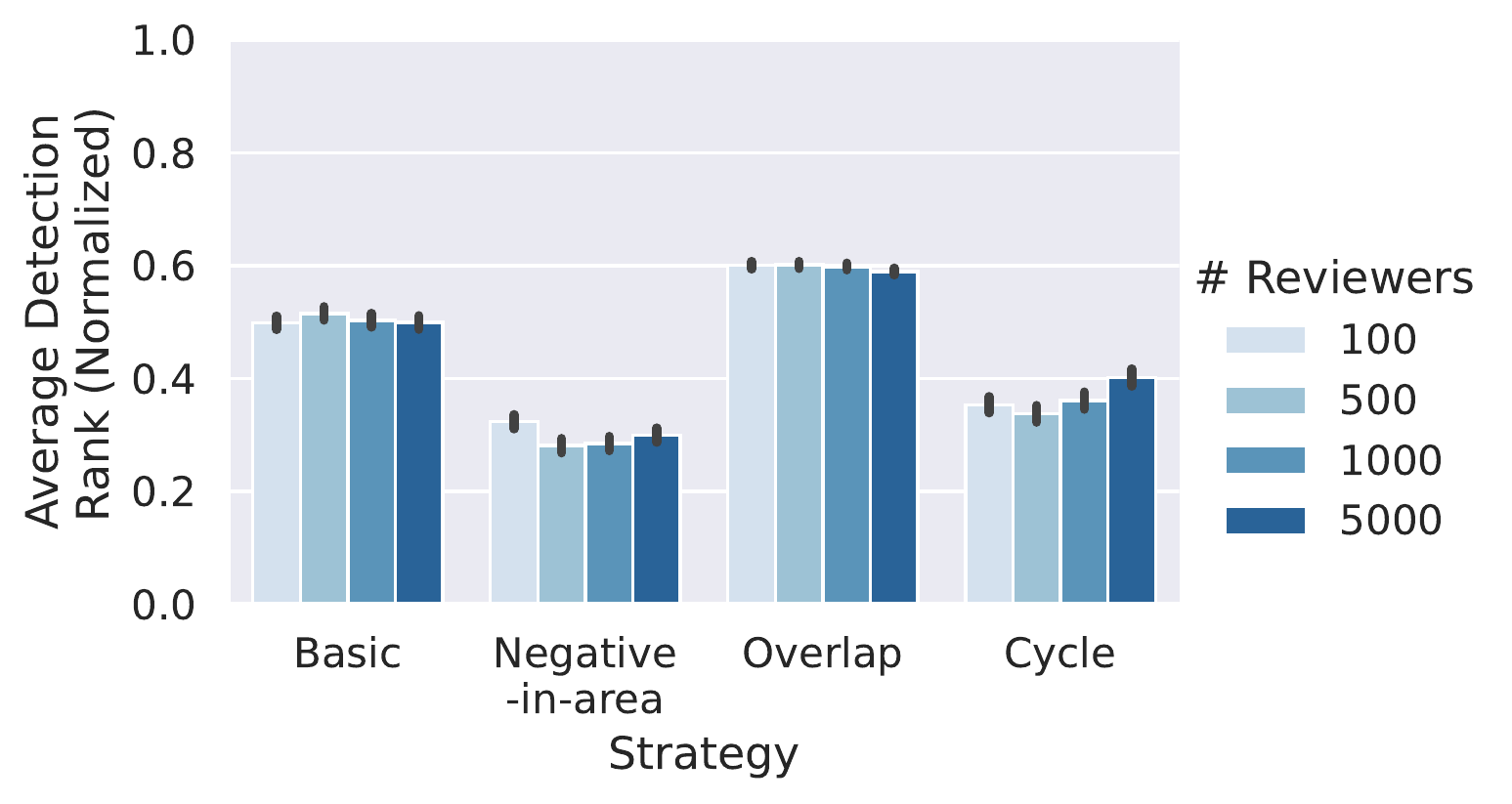}\caption{Counting detection}\label{fig:rank_counting_synth_n} \end{subfigure} \\
    \begin{subfigure}{0.48\textwidth}\includegraphics[width=1\textwidth]{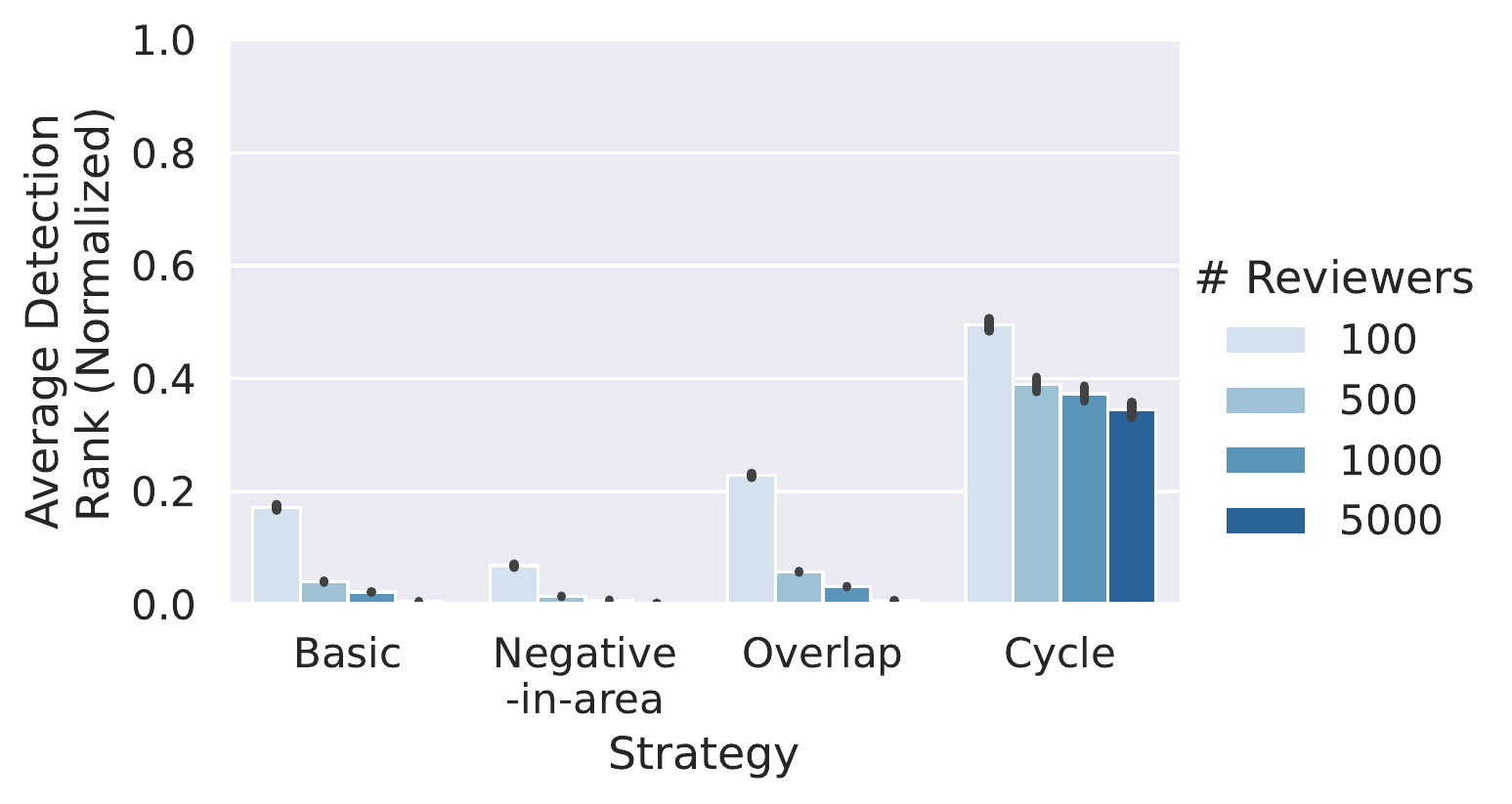}\caption{Ring detection}\label{fig:rank_cluster_synth_n} \end{subfigure} 
    \begin{subfigure}{0.48\textwidth}\includegraphics[width=1\textwidth]{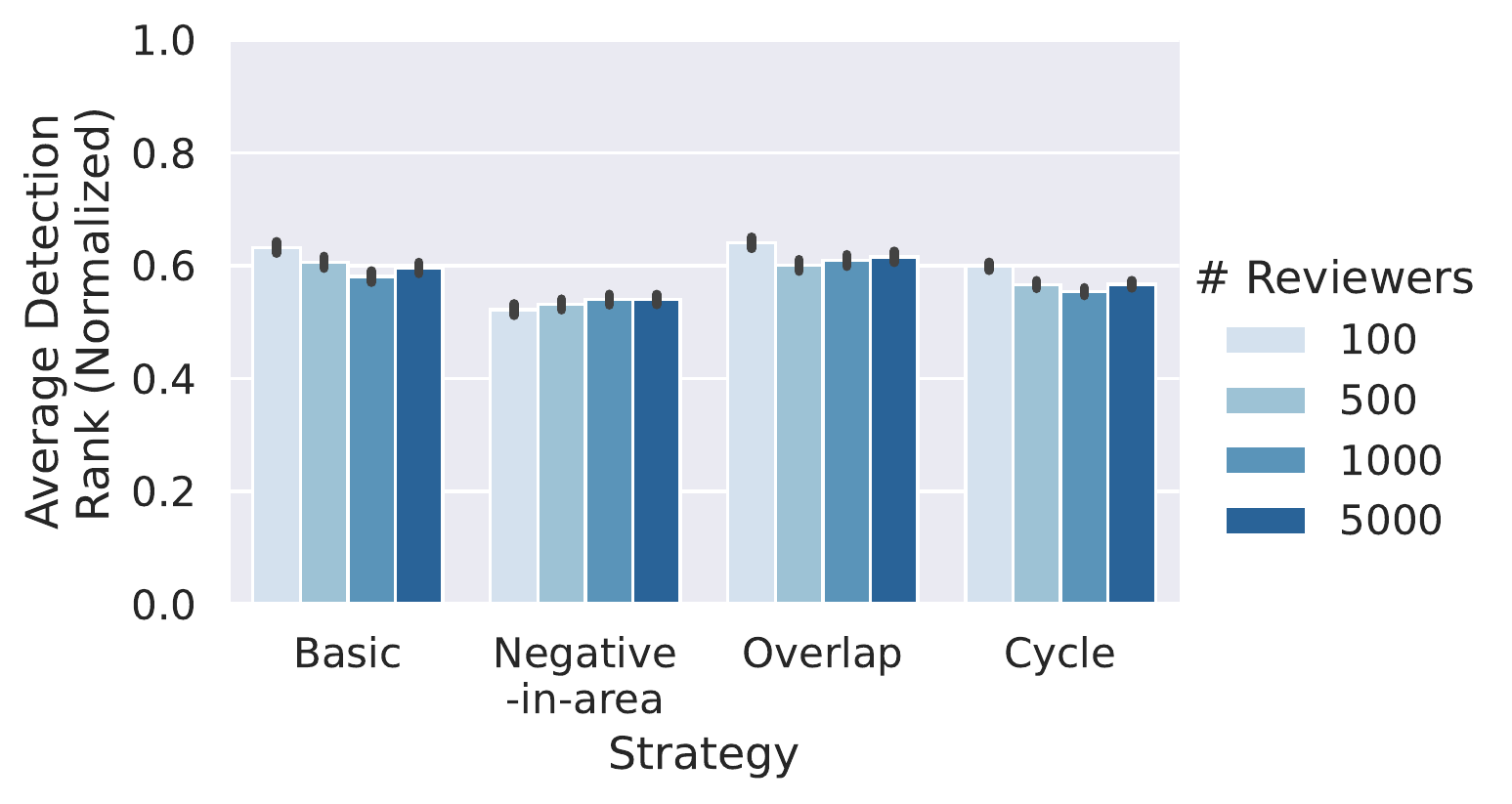}\caption{Low-rank detection}\label{fig:rank_lowrank_synth_n} \end{subfigure} 
    \caption{Results from synthetic scaled-up experiments with a malicious group size of 4. Different colors indicate different numbers of reviewers/papers.  
    In Figures~\ref{fig:rank_counting_synth_n}-\ref{fig:rank_lowrank_synth_n}, a lower normalized rank indicates the malicious reviewer was detected as more suspicious.  } 
    \label{fig:synth_n} 
\end{figure*}
}

\section{Participant Instructions} \label{apdx:instruct}
The participants were first verbally told about the problem of bid manipulation, along with a brief description of the activity. They were also told that participation is optional and ungraded. Some motivations to participate were stated: to get a hands-on experience in game-theoretic thinking, to help the community understand what kinds of bidding manipulation may be possible, and to experience what may be a fun exercise. 

The participants were subsequently provided written instructions, reproduced below. 
The text in brackets differed between participants. 

\textbf{Before beginning the activity:}
{\it 
As mentioned in class, we are running a fun game to give you a hands-on experience with game theory. This fun game is about strategic behavior in paper bidding for an academic conference. The activity is optional and won’t affect your grade in any way.

You (along with your classmates) will play the role of a reviewer for a fictional conference called FAIC (the Fake AI Conference). To determine which papers you should review, FAIC is asking you to “bid” on various papers. See the slides for more details. The activity has two parts: 

Part 1: You play the role of an honest reviewer and submit bids at FAIC according to your interests. Complete this part using [this personalized link].

Part 2: You play the role of a manipulative reviewer who wants to manipulate the assignment algorithm. You are working with a group of friends to do these manipulations: [group emails]. For this part, please read the instructions, discuss your strategy with your group, and then return to complete the activity. View this part using [this personalized link].
}

\textbf{Before the first phase (honest reviewing):}
{\it 
In this activity, you (along with your classmates) will be playing the role of a reviewer for a fictional conference called FAIC (the Fake AI Conference). FAIC is currently attempting to determine which papers each reviewer should be assigned to review based on their expertise and interests.

As a reviewer, your expertise is in the areas of [subject areas]. This means that you will be more likely to be assigned to papers matching this description. You are also an author on the paper [paper title] which you have submitted to FAIC. 

In order to further determine which papers you should be assigned to review, FAIC is asking you for your level of interest in each paper, commonly known as “bidding”. FAIC will then take the bids into account when assigning papers to reviewers.

Suppose that you are an honest reviewer at FAIC. This means that you should bid on papers according to your own personal interests, as if you were actually going to review the assigned papers. Keep in mind that each paper will be assigned 3 reviewers, and each reviewer will be assigned to at most 3 papers.
}

\textbf{Before the second phase (malicious reviewing):}
{\it 
Now, you will take the role of a malicious reviewer at FAIC. Such malicious reviewers work in groups with their friends with the goal of getting assigned to each other’s papers. Here is an example strategy that a pair of malicious reviewers working together might use: [image depicting two reviewers bidding positively on each other’s paper and negatively on all others].

The program chairs (PCs) organizing FAIC are aware that such manipulations can occur. If they notice any reviewers bidding suspiciously, they can manually modify the assignment to their liking. For example, the PCs may look through the bids to notice any reviewers that bid positively only on a single paper and choose to ignore those bids [image with example of such a malicious reviewer being detected]. As a malicious reviewer, you should be aware that your bidding manipulation may be detected.

To improve your paper’s chances of acceptance, you are working with your friends who have authored the papers [paper names]. Recall that you are an author on the paper [paper title] which you have submitted to FAIC. All of you are experts in [subject area]. 

Your goal is to strategically coordinate with your group as a team to bid so that you are assigned to review each other’s papers. You should communicate with them to discuss your bidding strategy (you can leave and return to this page at any time). Keep in mind that each paper will be assigned 3 reviewers, and each reviewer will be assigned to at most 3 papers. The more reviewers from within your group assigned to each paper, the higher that paper’s chances of acceptance are (which is good for your group).
}

\fullver{
\begin{figure*}[t!] 
    \centering
    \begin{subfigure}{0.48\textwidth}\includegraphics[width=1\textwidth]{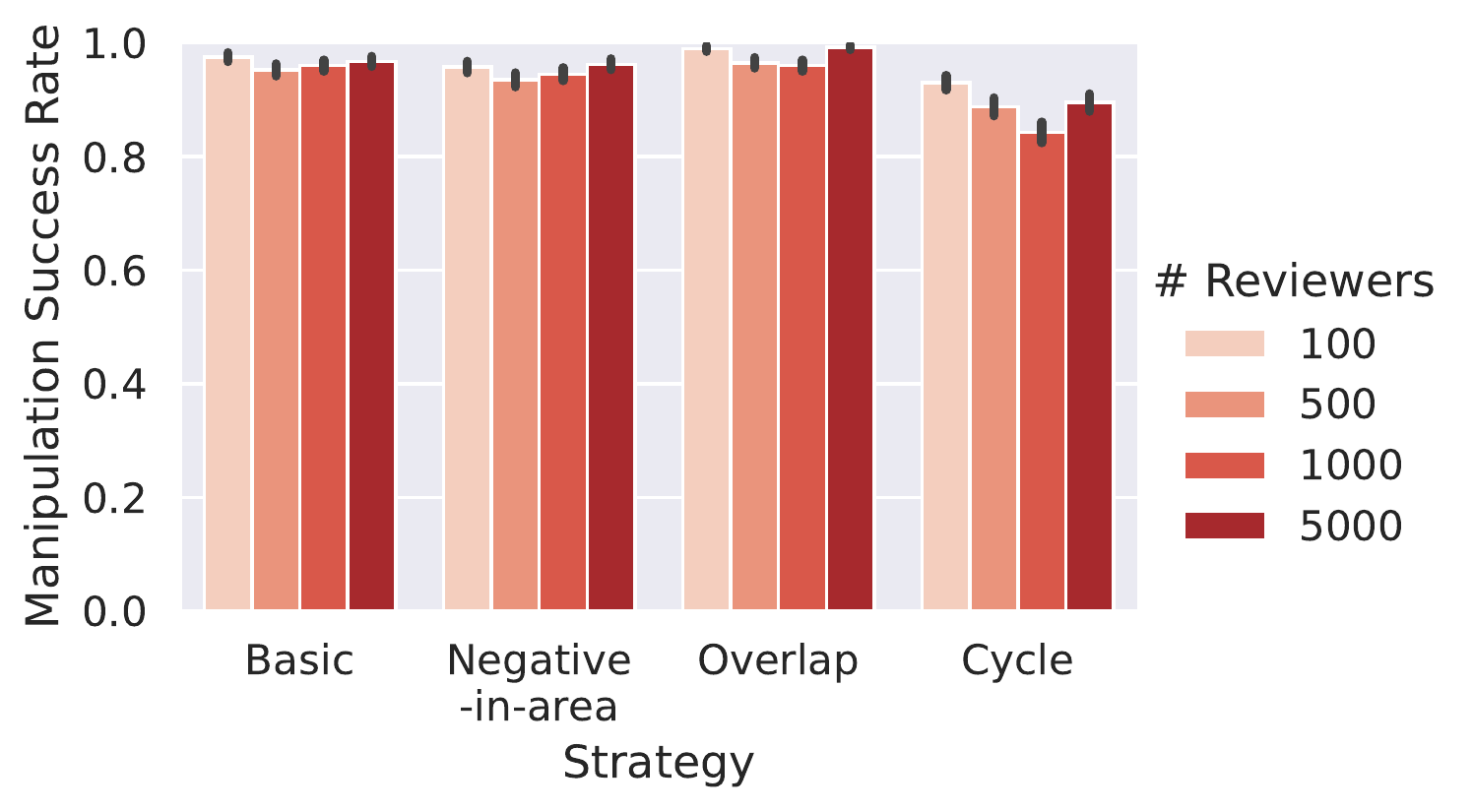}\caption{Malicious reviewer success rate}\label{fig:success_synth_n} \end{subfigure} 
    \begin{subfigure}{0.48\textwidth}\includegraphics[width=1\textwidth]{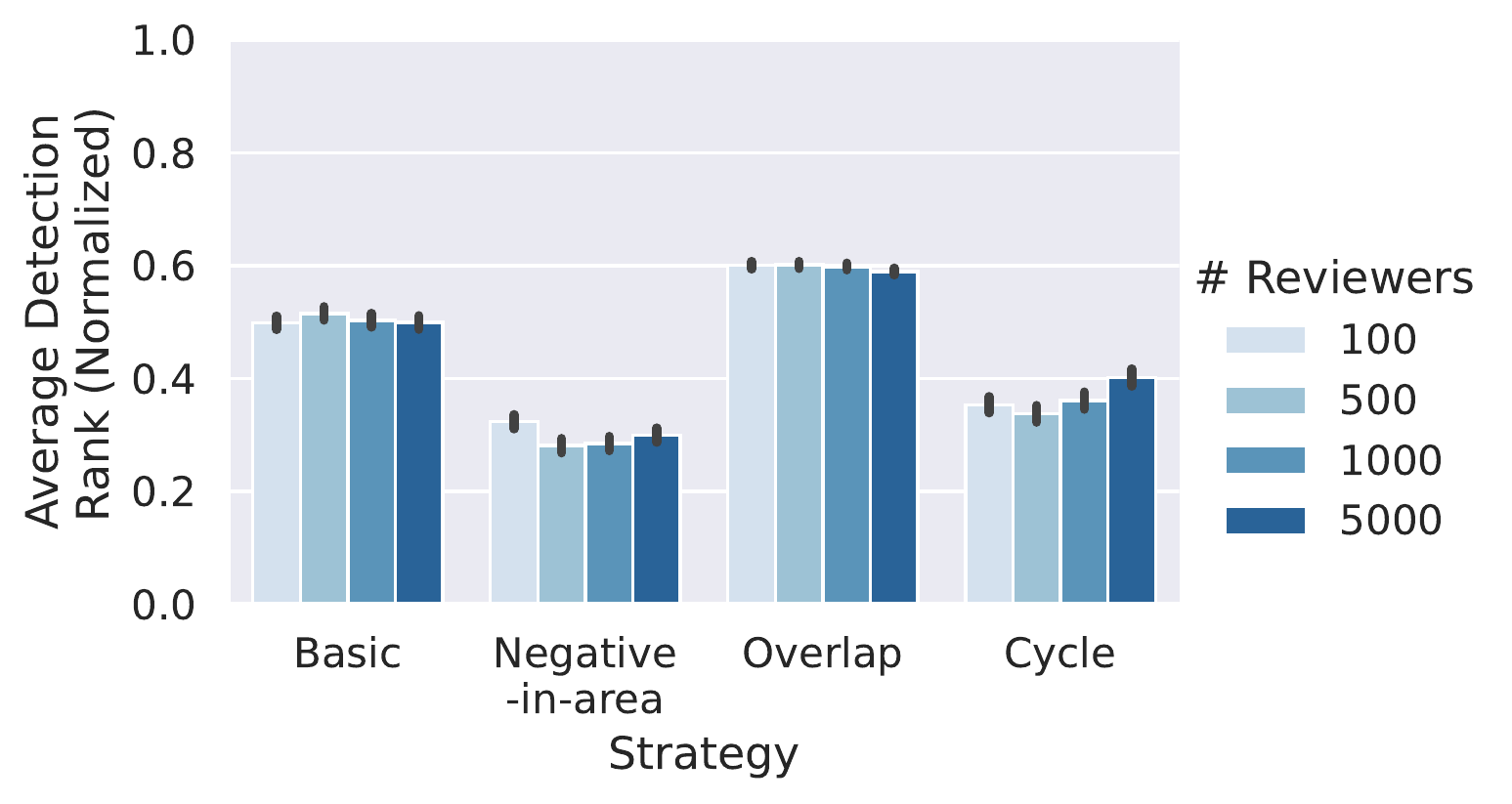}\caption{Counting detection}\label{fig:rank_counting_synth_n} \end{subfigure} \\
    \begin{subfigure}{0.48\textwidth}\includegraphics[width=1\textwidth]{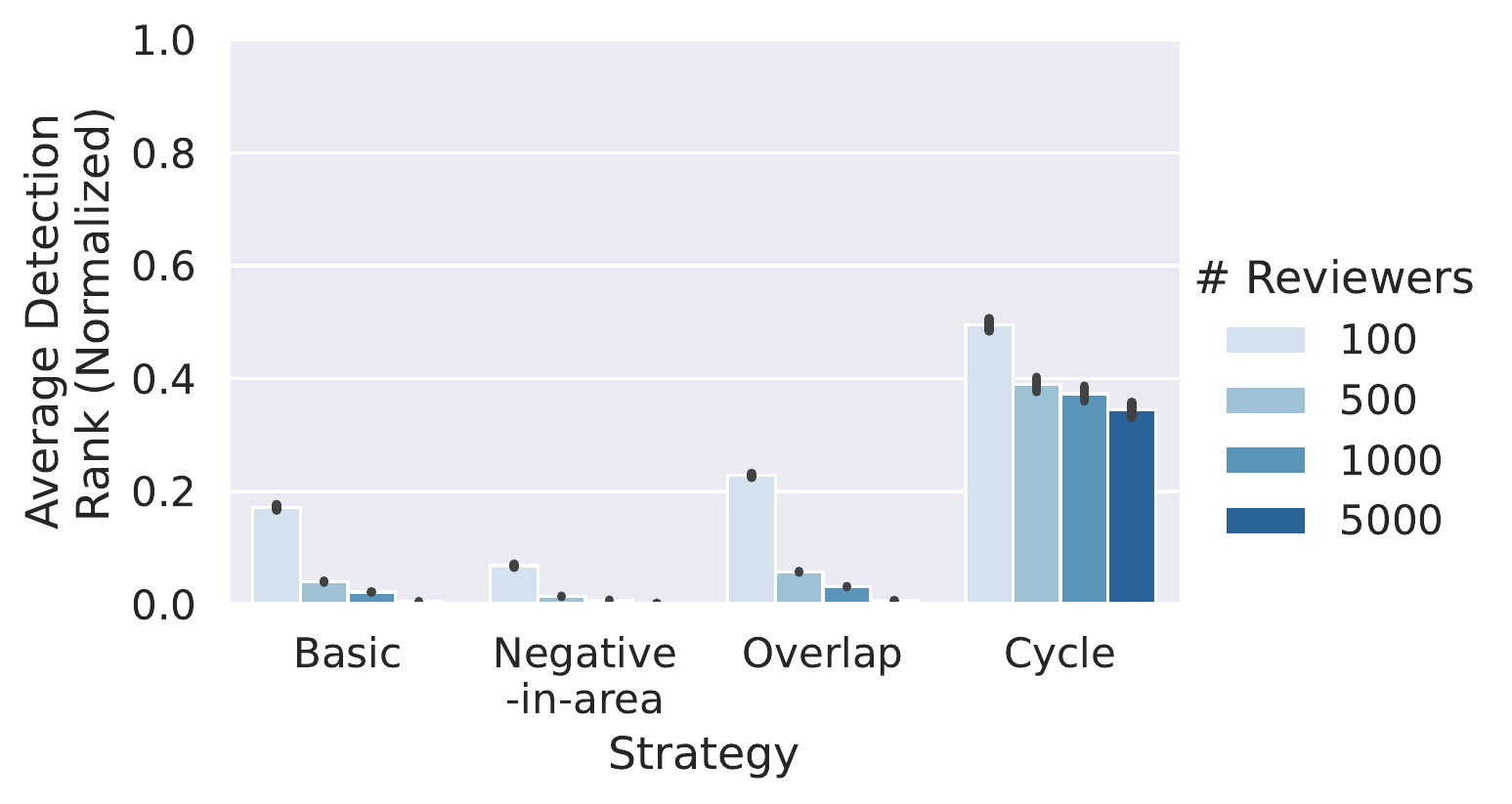}\caption{Ring detection}\label{fig:rank_cluster_synth_n} \end{subfigure} 
    \begin{subfigure}{0.48\textwidth}\includegraphics[width=1\textwidth]{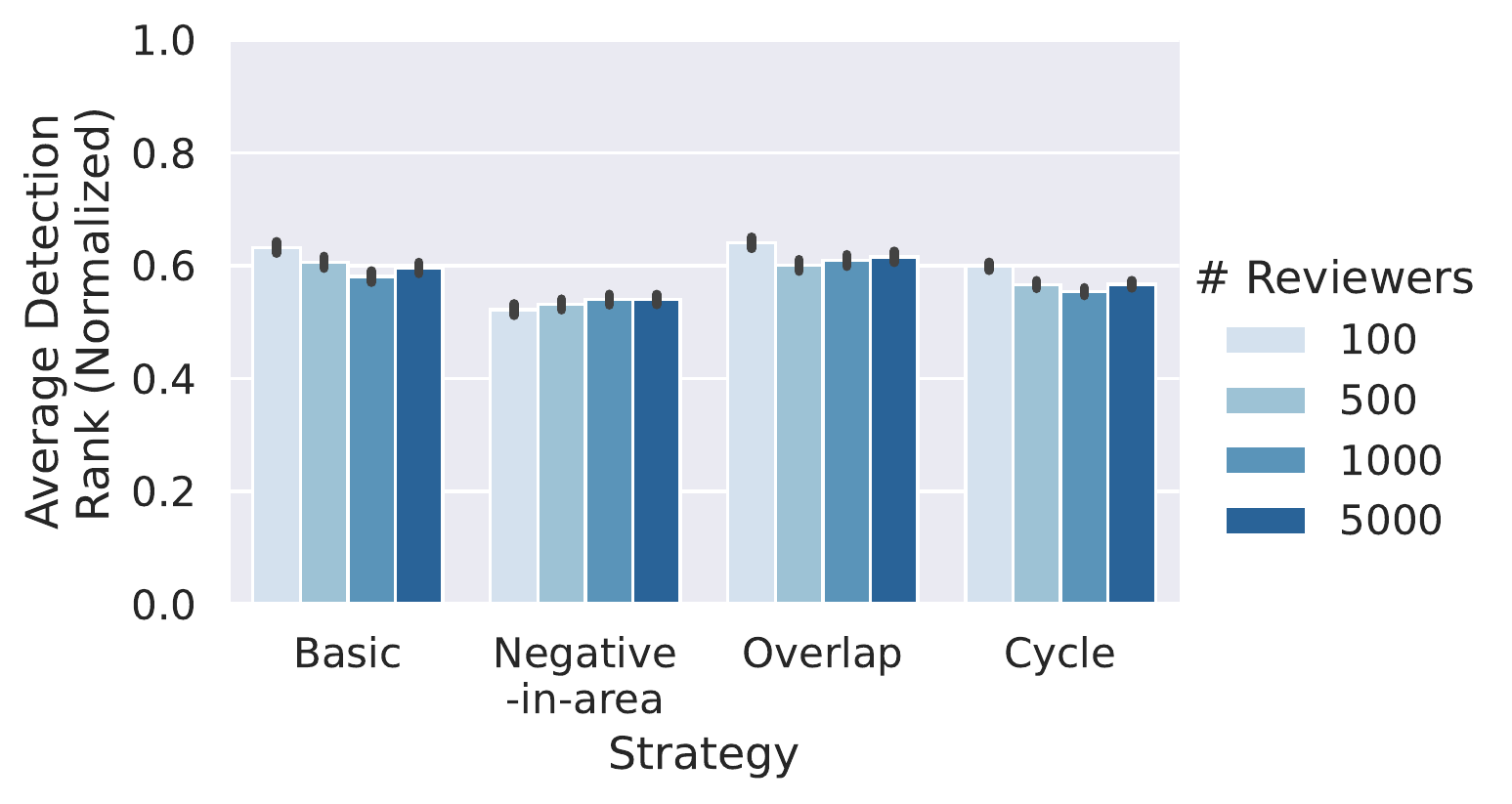}\caption{Low-rank detection}\label{fig:rank_lowrank_synth_n} \end{subfigure} 
    \caption{Results from synthetic scaled-up experiments with a malicious group size of 4. Different colors indicate different numbers of reviewers and papers.  
    In Figures~\ref{fig:rank_counting_synth_n}-\ref{fig:rank_lowrank_synth_n}, a lower normalized rank indicates the malicious reviewer was detected as more suspicious.  } 
    \label{fig:synth_n} 
\end{figure*}
}

\section{Additional Synthetic Results} \label{apdx:addl_results}
In Figure~\ref{fig:synth_n}, we display the results of additional scaled-up experiments (described in Section~\ref{sec:synth}). In these experiments, we fix the malicious group size at $4$ and vary the total number of reviewers and papers (held equal) between $100$ and $5000$. In Figures~\ref{fig:rank_counting_synth_n}-\ref{fig:rank_lowrank_synth_n}, the rank values (on the y-axis) are normalized by the number of reviewers so that they range from $0$ (most suspicious) to $1$ (least suspicious). These results show generally that within this range, the performance of the malicious reviewer strategies and detection algorithms is not affected by the size of the data. One exception is that in Figure~\ref{fig:rank_cluster_synth_n}, we see that the \textit{Ring Detection} algorithm does better as the number of reviewers and papers increases. This fits with the intuition that the detected rings stand out more among larger numbers of honest reviewers, which are unlikely to form rings.

\end{document}